\documentclass[10pt,
  aps,prl,twocolumn,preprintnumbers,superscriptaddress,
  nofootinbib,floatfix]{revtex4-2}

\usepackage[utf8]{inputenc}
\usepackage{amsmath}
\usepackage{bm}
\usepackage{amsfonts}
\usepackage{graphicx}
\usepackage{epstopdf}
\usepackage{hyperref}
\usepackage{array}
\usepackage{soul}
\usepackage{xcolor}
\usepackage{comment}
\usepackage{physics}
\usepackage{cancel}
\usepackage{slashed}

\enlargethispage{\baselineskip}
\usepackage{orcidlink}
\usepackage{csquotes}
\hypersetup{
     colorlinks   = true,
     citecolor    = black,
     breaklinks = true,
     urlcolor = black,
     linkcolor = black
}

\newcommand{\Appendix}{Appendix}

\newcommand{\rmi}[1]{{\mbox{\scriptsize #1}}}
\newcommand{\rmii}[1]{{\mbox{\tiny\rm{#1}}}}

\newcommand{\Mpl}{M_\rmii{Pl}}
\newcommand{\fphi}{f_\phi}
\newcommand{\mphi}{m_\phi}

\newcommand{\UPQ}{U(1)_\rmii{PQ}}
\newcommand{\OmegaPhiToday}{\Omega_{\phi,0}}
\newcommand{\Trel}{T_\rmi{rel}}
\newcommand{\thetarel}{\theta_\rmi{rel}}
\newcommand{\Hrel}{H_\rmi{rel}}
\newcommand{\Hosc}{H_\rmi{osc}}
\newcommand{\Tosc}{T_\rmi{osc}}
\newcommand{\aosc}{a_\rmi{osc}}
\newcommand{\gosc}{g_\rmi{osc}}
\newcommand{\Tmisosc}{T_\rmi{osc}^\rmi{mis}}
\newcommand{\LambdaPQ}{\Lambda_{\cancel{\rmii{PQ}}}}

\begin{document}
\title{Trapped and Unstable: Axion-like particle fragmentation at finite temperature}

\preprint{SISSA 15/2025/FISI}

\author{Nicklas Ramberg\,\orcidlink{0000-0003-1551-9860}\,}
\email{nramberg@sissa.it}
\affiliation{SISSA International School for Advanced Studies, Via Bonomea 265, 34136, Trieste Italy}
\affiliation{INFN Sezione di Trieste, Via Bonomea 265, 34136, Trieste Italy}
\affiliation{IFPU, Institute for Fundamental Physics of the Universe, Via Beirut 2, 34014 Trieste, Italy
}

\author{Daniel Schmitt\,\orcidlink{0000-0003-3369-2253}\,}
\email{dschmitt@itp.uni-frankfurt.de}
\affiliation{Institute for Theoretical Physics, Goethe University, 60438 Frankfurt am Main, Germany}
\date{\today}

\begin{abstract}
\noindent
We investigate the emergence of a resonant behavior in axion-trapped misalignment models featuring finite-temperature potential barriers.
As the temperature decreases and the field is released from its trapped configuration, inhomogeneities are exponentially amplified through an instability in their equation of motion, leading to the fragmentation of the axion field.
We show that this process constitutes a novel source of gravitational waves (GWs), analogous to those generated in zero-temperature axion fragmentation, but with distinct characteristics. 
We quantify the resulting GW spectrum, identifying the peak frequency and amplitude associated with the inhomogeneous axion dynamics. 
Our results indicate that the GW signal can be enhanced by up to two orders of magnitude compared to the standard fragmentation scenario, while exhibiting a markedly different spectral shape. 
The parameter space featuring both strong GW signals and reproducing the correct dark matter abundance is, however, limited.
\end{abstract}

\maketitle

\section{Introduction}
The QCD axion was originally introduced to resolve the strong CP problem in quantum chromodynamics (QCD) through the Peccei–Quinn (PQ) mechanism~\cite{Peccei:1977hh,Wilczek:1977pj,Weinberg:1977ma}. Beyond this motivation, the axion also stands out as a compelling candidate for cold dark matter (CDM), produced non-thermally through the misalignment mechanism~\cite{Preskill:1982cy,Abbott:1982af,Dine:1982ah}. 
In so-called trapped misalignment models, the axion potential acquires additional contributions in the form of explicit PQ-breaking or temperature-dependent operators, generating additional minima separated by a barrier~\cite{Nakagawa:2020zjr,DiLuzio:2021gos,DiLuzio:2021pxd,DiLuzio:2024fyt}.
For exceptionally light axions in the pre-inflationary scenario, ultraviolet (UV) completions based on discrete $\mathcal{Z}_{N}$ symmetries have been proposed as a natural realization of this framework~\cite{Hook:2018jle}.
Meanwhile, for post-inflationary PQ breaking, the QCD axion model is particularly predictive, leaving only a narrow window of viable parameter space~\cite{Gorghetto:2020qws,Benabou:2024msj,Saikawa:2024bta,Correia:2024cpk}. 

More generally, pseudoscalar fields arising as pseudo–Nambu–Goldstone bosons of broken global symmetries are referred to as axion-like particles (ALPs).
Such fields appear abundantly in string theory compactifications~\cite{Svrcek:2006yi,Arvanitaki:2009fg,Marsh:2015xka,Kim:1986ax} and have also been considered as potential inflaton candidates~\cite{Freese:1990rb,Pajer:2013fsa,Adshead:2015pva,Daido:2017wwb,Takahashi:2021tff,Domcke:2019qmm,Kitajima:2021bjq}. In this work, we focus on pre-inflationary ALPs as the underlying field content, while noting that the mechanism we propose may also find realization within QCD axion models.

In particular, we study ALPs that are initially trapped by a thermal barrier~\cite{DiLuzio:2024fyt,Gerlach:2025fkr}.
Then, the ALP remains in a metastable minimum until the barrier disappears, after which it rolls towards the true vacuum.
We show that the delayed onset of oscillations triggers a resonance in the ALP equation of motion, leading to a copious production of ALP quanta in the early Universe, resulting in the fragmentation of the ALP field~\cite{Fonseca:2019ypl,Chatrchyan:2020pzh,Morgante:2021bks,Madge:2021abk,Eroncel:2022efc,Eroncel:2022vjg,Chatrchyan:2023cmz}.\footnote{Note that such dynamics are absent in the standard misalignment mechanism, unless the initial misalignment angle is tuned close to the potential maximum~\cite{Greene:1998pb}.} 
Intriguingly, the generation of long-wavelength fluctuations generates anisotropies in the energy-momentum tensor of the cosmic fluid that act as a source of stochastic GWs~\cite{Kofman:1994rk,Greene:1997fu,Kofman:1997yn,Dufaux:2007pt,Dufaux:2008dn,Cook:2011hg,Buchmuller:2013lra,Figueroa:2017vfa,Figueroa:2020rrl,Bea:2021zol,Schmitt:2024pby,Dutka:2025oqt}.

A stochastic gravitational wave background~(SGWB) generated in the pre-nucleosynthesis universe propagates almost entirely undisturbed until today, hence offering deep insights into the physics of the very early universe, if ever observed~\cite{Caprini:2018mtu}.
While typical sources include, for example, first-order phase transitions~\cite{Kamionkowski:1993fg,Hindmarsh:2013xza,Schwaller:2015tja,Caprini:2019egz,Croon:2020cgk, Morgante:2022zvc,Kierkla:2022odc,Sagunski:2023ynd,Kierkla:2023von,Kierkla:2025qyz,Kierkla:2025vwp}, ALPs are attracting growing interest as a source of GWs~\cite{Anber:2012du,Domcke:2016bkh,Maleknejad:2016qjz,Cuissa:2018oiw,Machado:2018nqk,Machado:2019xuc,Ratzinger:2020oct,Salehian:2020dsf,Namba:2020kij,Kite:2020uix,Kitajima:2020rpm,Banerjee:2021oeu,Madge:2023dxc,Su:2025nkl,Gerlach:2025fkr}.
In particular, GW emission from axion fragmentation has been explored in monodromy~\cite{Chatrchyan:2020pzh} and kinetic misalignment~\cite{Madge:2021abk} scenarios. 
In this work, we show that trapped misalignment provides initial conditions that lead to ALP fragmentation and estimate the GW signal both analytically and numerically.
Remarkably, initial trapping significantly enhances the GW amplitude compared to previous works.
We further compute the ALP relic abundance, finding that simultaneously generating large GW amplitudes and the correct CDM abundance is hard to realize. 

We structure this paper as follows. 
First, we outline the model and calculate the analytic conditions for the release temperature and angle. 
Then, we identify the instability band, derive the conditions for efficient growth, and perform a numerical analysis of the resonant behavior. 
We compute the GW spectrum for different benchmarks and discuss its characteristics, demonstrating that trapped misalignment offers a viable GW source.

\section{Thermally trapped misalignment}
In the ordinary misalignment mechanism, one considers an ALP~$\phi$ with a potential
\begin{equation}
    V(\phi) = \mphi^2 \fphi^2 \left(1 - \cos\left(\frac{\phi}{f}\right)\right) \, .
\end{equation}
Initially displaced by the misalignment angle $\theta_i \sim \mathcal{O}(1)$, the ALP starts to oscillate around the potential minimum as the Hubble parameter drops below the ALP mass $H\sim \mphi$.
This corresponds to the temperature
\begin{align}\label{eq:T_osc,aa}
    \Tmisosc = \left(\frac{90}{\pi^2 g_\epsilon}\right)^\frac{1}{4} \sqrt{\mphi \Mpl}\,.
\end{align}
Here, $g_\epsilon$ denotes the effective relativistic degrees of freedom and $\Mpl$ is the reduced Planck mass.
Subsequently, the ALP energy density follows a matter-like scaling $\rho_\phi \propto a^{-3}$, making the pseudoscalar a viable CDM candidate.

\begin{figure}
    \centering
    \includegraphics[width=\linewidth]{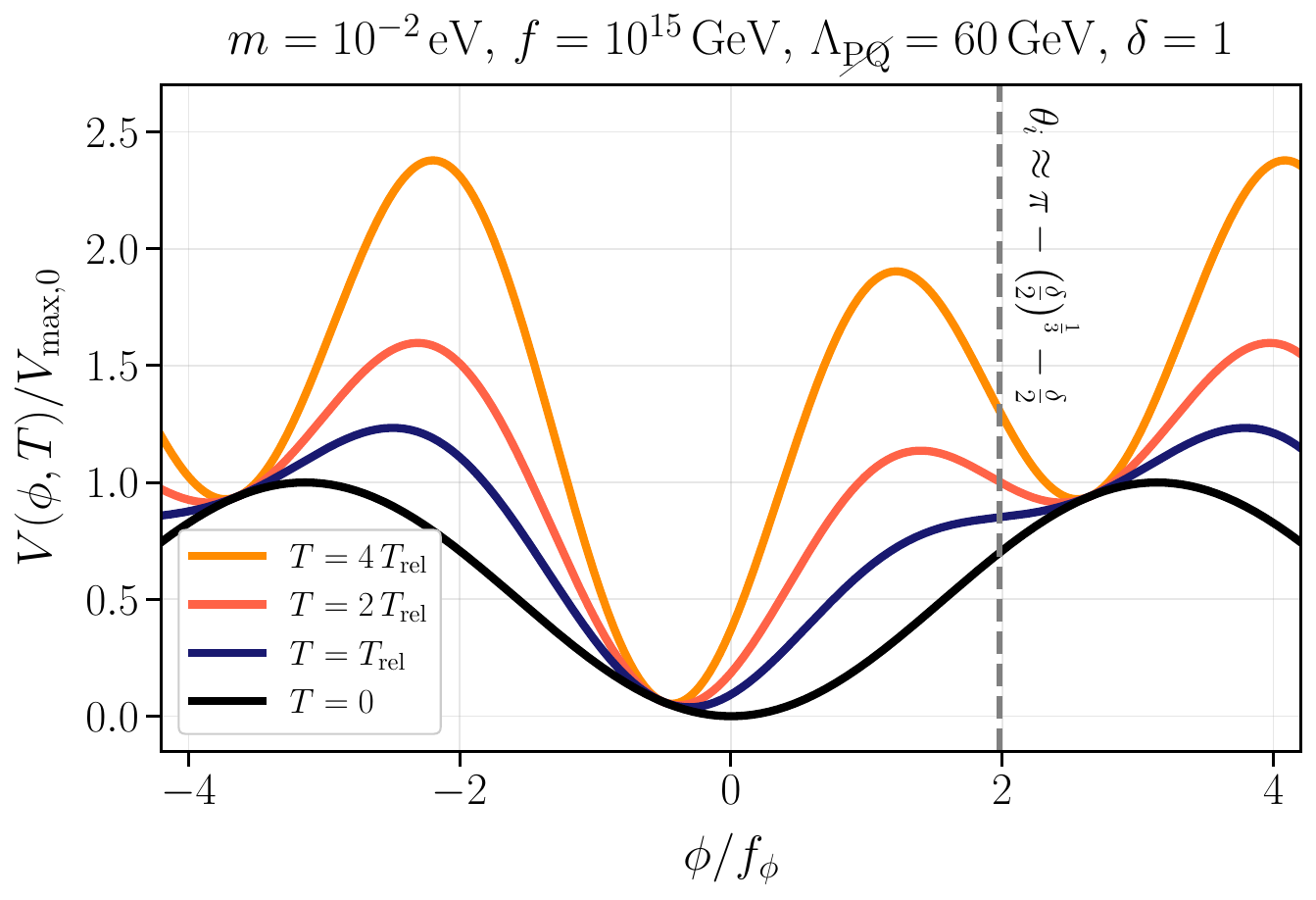}
    \caption{Exemplary ALP potential~\eqref{eq:trapping_potential}, employing the indicated benchmark parameters. Initially, a thermal barrier prevents the ALP from rolling. As the barrier vanishes around~$\Trel$, the false minimum becomes a saddle point and oscillations about the true minimum start. The  numerical evaluation of the finite-temperature release angle~(dashed gray) agrees well with our approximate solution~\eqref{eq:theta_rel}.}
    \label{fig:potential}
\end{figure}

Trapped misalignment has been suggested as an alternative mechanism for ALP production in the early Universe~\cite{DiLuzio:2021gos,DiLuzio:2021pxd,DiLuzio:2024fyt}.
Here, further explicit PQ-breaking terms alter the structure of the ALP potential, generating additional minima.
This can trap the ALP in a false minimum, thereby delaying the onset of ALP oscillations.
In this work, we consider the temperature-dependent trapping potential from~\cite{DiLuzio:2024fyt,Gerlach:2025fkr}
\begin{equation}\label{eq:trapping_potential}
\begin{split}
    V(\phi,T) = &\mphi^2 \fphi^2 \left(1 - \cos\left(\frac{\phi}{\fphi}\right)\right) \\
    &+ \LambdaPQ^{4-q} T^q \left(1 - \cos\left(n \frac{\phi}{\fphi} + \delta\right)\right)\,, 
\end{split}
\end{equation}   
where $m_{\phi}$ is the axion mass, $f_{\phi}$ its decay constant, $\LambdaPQ$ is the energy scale at which explicit PQ breaking operators emerge, $n$ denotes the multiplicity of minima, and $\delta$ is the phase shift. 
For simplicity, we focus on $n=2$ hereafter.\footnote{Note that it is straightforward to extend our results to an arbitrary number of additional minima.}
An exemplary ALP potential is shown in fig.~\ref{fig:potential}.
At high temperature, the $T$-dependent terms in the potential~\eqref{eq:trapping_potential} generate a false minimum that vanishes at the release temperature $T = \Trel$.
The time when the field is released from the trapping potential is defined by the following conditions: 
\begin{equation}\label{eq:trapping_conditions_general}
    \frac{\partial V(\phi,T)}{\partial \phi} = 0 \, , \quad \frac{\partial^2 V(\phi,T)}{\partial \phi^2} = 0 \, .
\end{equation}
This corresponds to a saddle point in the potential.
In the following, we analytically derive approximate conditions on the model parameters for trapping to terminate.

First, from eq.~\eqref{eq:trapping_conditions_general} we find a relation for the position of the false minimum, i.e., the release angle
 \begin{equation}
    \tan(\thetarel) = \frac{1}{n}\tan(n \thetarel + \delta)\,.
\end{equation}
In the case of $n=2$ one can reformulate the equation into an algebraic trigonometric equation of cubic order 
\begin{equation}\label{eq:release_condition_trig}
    2\tan^3(\thetarel) + 3\tan(\delta)\tan^2(\thetarel) + \tan(\delta) = 0 \,.
\end{equation}
If $\delta=0$, the equation becomes a triple-root.  
For small but non-zero $\delta$ one can Taylor-expand $\tan(\delta)~\approx~\delta~+~\mathcal{O}(\delta^3)$, which reduces eq.~\eqref{eq:release_condition_trig} to 
\begin{equation}
            2t^3 + 3\delta t^2 + \delta = 0\,,
\end{equation}
with $t = \tan(\thetarel)$. 
We will only concern ourselves with real-valued solutions to this equation, which can be shown to have the form 
\begin{equation}
    \thetarel \simeq \arctan\left(-2^{-\frac{1}{3}}\delta^{\frac{1}{3}} - \frac{\delta}{2} + \mathcal{O}(\delta^{\frac{5}{3}})\right) + k\pi\,,\quad k\in \mathbb{Z}\,,
\end{equation}
where $k$ is an integer number.
This can be finally formulated as \begin{equation}\label{eq:theta_rel}
    \thetarel \simeq -2^{-\frac{1}{3}}\delta^{\frac{1}{3}} - \frac{\delta}{2} +\mathcal{O}(\delta^{\frac{5}{3}}) + k\pi\,,\qquad k\in \mathbb{Z}\,.
\end{equation}
In the above expression, $(\delta/2)^{1/3}$ is the leading-order term, and the linear term is second-order.  
We have checked its validity at $\Trel$ and found that only using the first term in the expansion is sufficient to estimate the release temperature as a function of the initial angle. 
In full generality, one can solve eq.~\eqref{eq:release_condition_trig} for a certain value of $\delta$ and insert that value for $\thetarel$ to then find $\Trel$. 

We compute the release temperature by requiring that the ALP becomes massless at $\thetarel$ defined by eq.~\eqref{eq:theta_rel}.
That is, we solve the second trapping condition~(cf.~eq.~\eqref{eq:trapping_conditions_general}), which gives
\begin{equation}
    \Trel \approx \left(\frac{\mphi^2 \fphi^2 \cos(\thetarel)}{4 \LambdaPQ^{4-q} \cos(2\thetarel + \delta)}\right)^\frac{1}{q} \, .
\end{equation}

Finally, we need to check whether trapping actually takes place.
If, e.g., $\LambdaPQ$ is too small, we have $\Trel > \Tmisosc$, where the latter marks the onset of oscillations in the ordinary misalignment mechanism defined via $H=\mphi$; see eq.~\eqref{eq:T_osc,aa}.
%\begin{align}\label{eq:T_osc,aa}
%    \Tmisosc = \left(\frac{90}{\pi^2 g_\epsilon}\right)^\frac{1}{4} \sqrt{\mphi \Mpl}\,.
%\end{align}
%Here, $g_\epsilon$ denotes the effective relativistic degrees of freedom and $\Mpl$ is the reduced Planck mass.
Then, $\Hrel > \mphi$, and the ALP would remain overdamped until the temperature drops to $\Tmisosc$.
Therefore, we define the oscillation temperature as
\begin{equation}\label{eq:Tosc}
    T_\rmi{osc} = \mathrm{min}\{T_\rmi{rel}, T_\rmi{osc}^\rmi{mis}\} \, ,
\end{equation}
which is related to the Hubble parameter at the start of oscillations through
\begin{equation}\label{eq:Hosc}
    \Hosc = \left(\frac{\rho_r}{3\Mpl^2}\right)^\frac{1}{2} \, ,
\end{equation}
with $\rho_r = (\pi^2/30) \gosc \Tosc^4$ being the energy density of the radiation bath.

\begin{figure*}
    \centering
    \includegraphics[width=\linewidth]{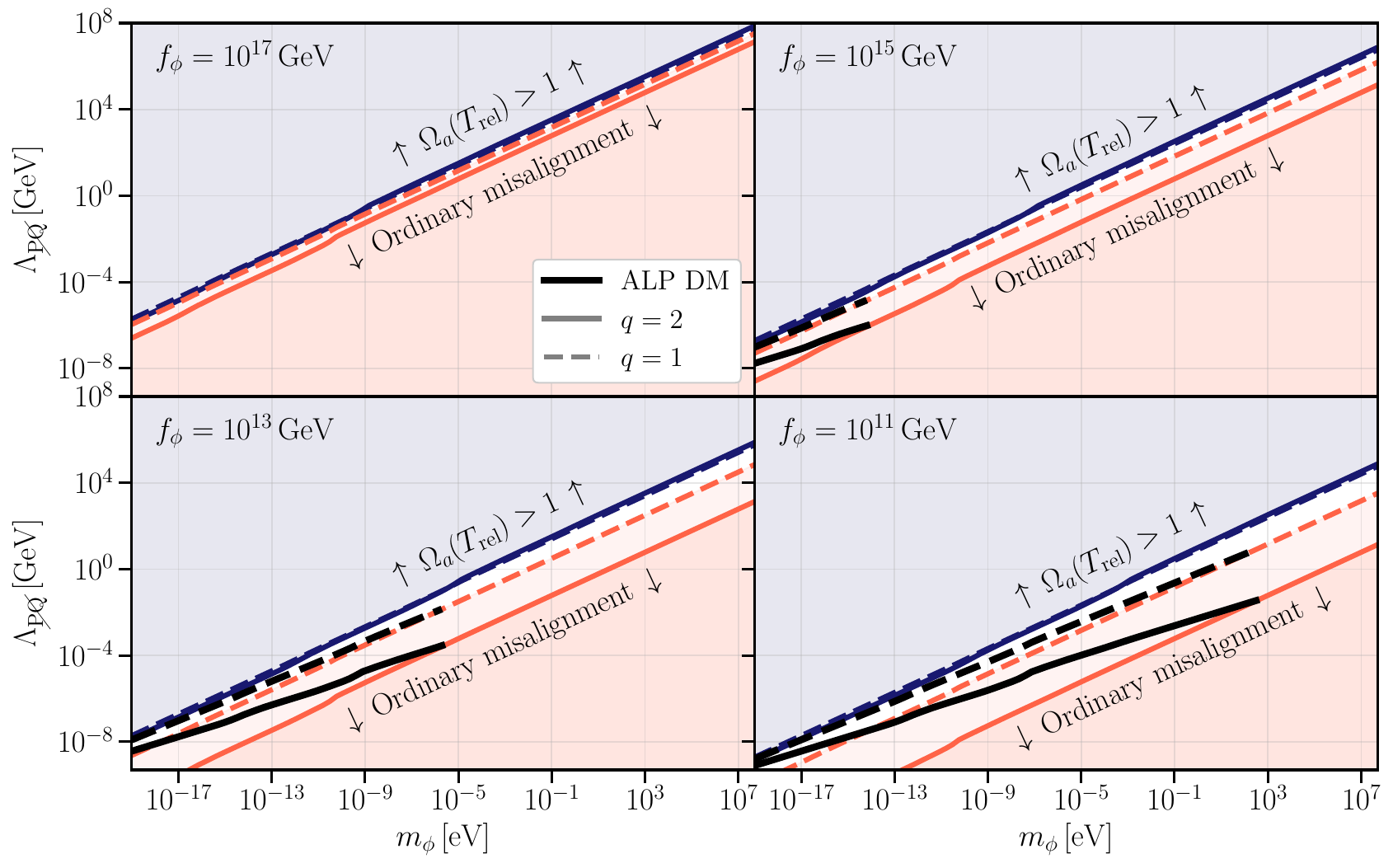}
    \caption{Viable parameter space for ALP fragmentation in the $\mphi-\LambdaPQ$ parameter space. Each panel corresponds to a different choice of the ALP decay constant, $\fphi \in \{10^{11},10^{13},10^{15},10^{17}\}\,\mathrm{GeV}$. Solid~(dashed) lines employ $q=2$~($q=1$), while the black lines indicate the parameter space where the correct CDM abundance is produced. In the blue-shaded region, the trapping period becomes too extended such that the ALP overcloses the Universe at the onset of oscillations. In the orange-shaded regime, the setup reduced to ordinary misalignment, where fragmentation is inefficient~(cf.~eq.~\eqref{eq:m_over_H_est}). Generally, smaller $\fphi$ and larger $q$ open up the viable parameter in terms of $\LambdaPQ$. In addition, large decay constants severely limit the mass range where the ALP can be CDM.}
    \label{fig:overview_parameter_space}
\end{figure*}

\section{ALP fragmentation}
In this section, we outline the dynamics of the resonant behavior leading to the fragmentation of the ALP field, derive the conditions for efficient growth of fluctuations, and discuss the relic ALP abundance.
\subsection{Equations of motion}
The ALP equation of motion reads
\begin{equation}\label{eq:ALP_eom}
    \Ddot{\phi} + 3 H\Dot{\phi} - \frac{1}{a^2} \nabla^2 \phi + \frac{\partial V(\phi,T)}{\partial\phi} = 0 \, ,
\end{equation}
where dots indicate derivatives with respect to cosmic time and $a$ is the scale factor of the Universe.
In this work, we restrict ourselves to a perturbative approach to compute the ALP evolution.
That is, we decompose the pseudoscalar into a homogeneous mode and small fluctuations
\begin{equation}
    \begin{split}
        \phi(t,\boldsymbol{x}) &= \phi(t) + \delta\phi(t,\boldsymbol{x}) \\
        &=\phi(t) + \left(\int \frac{\mathrm{d}^3k}{(2\pi)^3} a_k u_k(t)\exp(i\boldsymbol{k}\boldsymbol{x}) + \mathrm{h.c.}\right) \, .
     \end{split}
\end{equation}
We introduce creation~($a_k$) and annihilation~($a_k^\dagger$) operators that satisfy the commutation relation~$[a_k,a_{k'}^\dagger]~=~(2\pi)^3~\delta^{(3)}(k-k')$.
The time evolution is governed by the mode functions $u_k(t)$, which we initialize in the Bunch-Davies vacuum.
For small fluctuations $\delta\phi \ll \phi$, we can expand the derivative of the potential as~\cite{Fonseca:2019ypl,Madge:2021abk}
\begin{equation}\label{eq:potential_expansion}
    \frac{\partial V}{\partial\phi} = \frac{\partial V}{\partial\phi} + \frac{\partial^2 V}{\partial\phi^2} \delta\phi + \frac{1}{2}\frac{\partial^3 V}{\partial\phi^3} \delta\phi^2 + ... \, .
\end{equation}
Inserting this expression into eq.~\eqref{eq:ALP_eom}, switching to Fourier space, and employing conformal time~$\tau$, we obtain two coupled equations of motion:
\begin{align}\label{eq:lin_eom_hom}
    \phi'' + 2 a H\phi' + a^2 \frac{\partial V}{\partial\phi} + \frac{1}{2} \frac{\partial^3 V}{\partial\phi^3} \int \frac{\mathrm{d}^3 k}{(2\pi)^3} |u_k|^2 &= 0 \,,\\\label{eq:lin_eom_fluc}
    u_k'' + 2aH u_k' + \left[k^2 + a^2 \frac{\partial^2 V}{\partial\phi^2}\right]u_k &=0\, .
\end{align}
Here, the last term in eq.~\eqref{eq:lin_eom_hom} governs the backreaction from the fluctuations onto the zero mode.

In the limit of a negligible Hubble rate, eq.~\eqref{eq:lin_eom_fluc} reduces to a Mathieu equation~\cite{McLachlan:1947:TA}, which exhibits a resonant behavior in certain momentum bands.
The dominant instability band is approximately given by~\cite{Fonseca:2019ypl} 
\begin{equation} \label{eq:instability_band}
    \frac{\dot{\phi}^2}{4\fphi^2} - \frac{\mphi^2}{2} \lesssim \left(\frac{k}{a}\right)^2 \lesssim \frac{\dot{\phi}^2}{4\fphi^2} + \frac{\mphi^2}{2} \, .
\end{equation}
For this estimate, we have neglected the finite-$T$ contributions to the potential, as they quickly become subdominant to the zero-$T$ part after the onset of ALP oscillations.
Hence, the mode that experiences the fastest growth is
\begin{equation}\label{eq:peak_momentum}
    \left(\frac{ k_\rmi{peak}}{a}\right)^2= \frac{\dot{\phi}^2}{4 \fphi^2} \approx \frac{\mphi^2}{2} (1-\cos(\theta_i))\, .
\end{equation}
In the second step, we have assumed that the initial ALP potential energy is converted into kinetic energy, again neglecting the finite-$T$ contribution to the ALP potential.
For $\thetarel \sim \mathcal{O}(1)$, the maximum growth rate reads~\cite{Kofman:1997yn,Fonseca:2019ypl}
\begin{equation}
    \frac{\omega_\rmi{max}}{a} \approx \frac{\mphi}{4}\, .
\end{equation}
Modes within the instability band are then exponentially amplified, 
\begin{equation}
    u_k \propto \exp(\sqrt{\omega_\rmi{max}^2 - (k-k_\rmi{peak})^2}\,\tau) \, ,
\end{equation}
eventually leading to the fragmentation of the ALP.
For a given set of input parameters, we solve eqs.~\eqref{eq:lin_eom_hom} and~\eqref{eq:lin_eom_fluc} numerically and extract the time evolution of the mode functions.
The results of a benchmark simulation can be found in the appendix.
In the following, we derive the necessary conditions for resonant growth to be efficient.

\subsection{Conditions for efficient growth}
To have ALP fragmentation, we need to ensure that ALP fluctuations occur in the first place.
This implies a bound on the initial misalignment angle, which is required to be within the concave part of ALP potential,
\begin{equation}\label{eq:condition_trapping}
    |\thetarel| > \frac{\pi}{2} \, .
\end{equation}
If $|\thetarel| < \pi/2$, the ALP would smoothly shift to the true minimum~\cite{DiLuzio:2024fyt}.
Via eq.~\eqref{eq:theta_rel}, this can be translated into a bound on the phase shift $\delta$ between the two contributions to the potential~\eqref{eq:trapping_potential}.

The energy density of the fluctuations grows as $\rho_{\delta\phi}\propto \exp(2|\omega|\tau)$. 
For amplification to be cosmologically efficient, the growth rate has to exceed the Hubble rate, i.e.,
\begin{equation}\label{eq:m_over_H_condition}
    m_\phi \gtrsim H \, .
\end{equation}
If this was not the case, the fluctuations would be overdamped~(cf.~eq.~\eqref{eq:lin_eom_fluc}).
In the ordinary misalignment mechanism, the onset of ALP oscillations is defined by $m_\phi~\sim~H$.
Then, amplification of unstable ALP modes can barely keep up with the expansion rate of the Universe.
Even if the initial misalignment angle fulfills eq.~\eqref{eq:condition_trapping}, cosmic expansion, $\dot{\phi}^2 \propto (a_\rmi{osc}/a)^3$, quickly shifts the instability band~\eqref{eq:instability_band} into the IR.
This shuts off the amplification of the initially unstable ALP modes, and ALP fragmentation becomes cosmologically inefficient.
Therefore, we exclude the region where the thermal barrier in eq.~\eqref{eq:trapping_potential} vanishes above $\Tmisosc$.

If trapping is effective, the Hubble parameter at the onset of oscillations decreases with respect to the conventional misalignment scenario.
Then, $\mphi > H_\rmi{rel}$ is always fulfilled, with ALP amplification becoming increasingly efficient as the trapping period is extended.
We are interested in the case where the exponential resonance can convert the entire initial energy density of the zero mode into fluctuations.
To obtain a rough estimate when this occurs, we again neglect the finite-$T$ contributions to the potential.
We employ $\rho_{\phi,\rmi{osc}} \sim \mphi^2 \fphi^2$  for the zero mode and 
\begin{equation}
    \rho_{\delta\phi} = \frac{1}{2a^2} \int \frac{\mathrm{d}^3k}{(2\pi)^3} \left(k^2 |u_k|^2 + |u_k'|^2\right) \approx \frac{k_\rmi{peak}^4}{\aosc^4 (4\pi)^2} \, ,
\end{equation}
to estimate the initial fluctuation energy density.
By imposing $k_\rmi{peak}/\aosc \sim \mphi$ and $\rho_{\delta\phi} \propto \exp(\mphi t/2)$, we can compute the elapsed cosmic time until $\rho_\delta\phi \sim \rho_{\phi,\rmi{osc}}$,
\begin{equation}
    t \sim 2\mphi^{-1} \ln\left(16\pi^2 \frac{\fphi^2}{\mphi^2}\right) \, ,
\end{equation}
translating to
\begin{equation}
    \frac{a_\star}{a_\rmi{osc}} \approx 1 + \frac{2\Hosc}{\mphi} \ln\left(16\pi^2 \frac{\fphi^2}{\mphi^2}\right) \, .
\end{equation}
To estimate the time when the resonant behavior shuts off, we compute the scale factor at which the largest initially unstable $k$-mode with $(k_\rmi{max}/\aosc)^2\sim 3\mphi^2/2$ is shifted out of the instability band. 
Since the fluctuations are produced predominantly with $k\lesssim \mphi$, we assume a matter-like scaling $k/a \propto \dot{\phi}^2 \propto a^{-3}$, such that
\begin{equation}
    \left(\frac{k_\rmi{max}}{\aosc}\right)^2 \left(\frac{\aosc}{a_\rmi{close}}\right)^3 = \frac{\mphi^2}{2} \, ,
\end{equation}
which gives
\begin{equation}
    \frac{a_\rmi{close}}{\aosc} \sim 3^\frac{1}{3} \, .
\end{equation}
Note that while this result neglects variations of the misalignment angle, it provides an informative criterion on the parameter space that allows for fragmentation.
We demand $a_\star < a_\rmi{close}$, which yields
\begin{equation}\label{eq:m_over_2H}
    \frac{\mphi}{2 \Hosc} \gtrsim  (3^{\frac{1}{3}} - 1)^{-1} \ln\left(16\pi^2 \frac{\fphi^2}{\mphi^2}\right) \, .
\end{equation}
Hence, for typical values of the ALP mass and decay constant, we find 
\begin{equation}\label{eq:m_over_H_est}
    \frac{\mphi}{\Hosc} \gtrsim \mathcal{O}(500) \, ,
\end{equation}
which we have also verified numerically.
This condition can then be translated to a lower bound on the duration of the trapping period, mainly controlled by $\LambdaPQ$.

For a consistent cosmological evolution, it is crucial to check that the ALP does not overclose the Universe at the onset of oscillations. 
Since we only consider the regime where the ALP is cosmologically stable, overclosure would necessarily lead to a matter-dominated Universe.
Therefore, we exclude the parameter space where
\begin{equation}
    \rho_\phi(T_\rmi{osc}) \approx m_\phi^2 f_\phi^2 (1- \cos\left(\theta_i\right)) > \frac{\pi^2}{30} g_\rmi{osc} T_\rmi{osc}^4 = \rho_{r}(T_\rmi{osc})\, ,
\end{equation}
with $\rho_r$ being the energy density of the primordial plasma.

An overview of the $m_\phi - \LambdaPQ$ parameter space is shown in fig.~\ref{fig:overview_parameter_space} for different decay constants $f_\phi$.
We employ $\delta = 0.01$; varying $\delta$ in the range $[0,\pi)$ has a negligible impact on the parameter space.
The blue- and orange-shaded regions display the regions where the ALP overcloses the Universe and where the setup reduces to ordinary misalignment, respectively. 
Here, the dashed (straight) line corresponds to $q=1$~($q=2$). 
In the region enclosed by the orange and blue lines, ALP fragmentation is possible, given eq.~\eqref{eq:m_over_2H} is fulfilled.
Note that for larger $q$, the range of possible $\UPQ$ symmetry-breaking scales~$\LambdaPQ$ increases.
This is because of the modified temperature dependence of the second term in eq.~\eqref{eq:trapping_potential}, requiring larger shifts of $\LambdaPQ$ to achieve a given release temperature.  

\subsection{Relic abundance}
In the following, we derive the relic ALP abundance $\OmegaPhiToday$ as a function of the model parameters, and identify the parameter space where the ALP can consitute all of CDM, i.e., $h^2\OmegaPhiToday = 0.12$~\cite{Planck:2018vyg}.

From the onset of oscillations, the ALP zero mode follows a matter-like scaling, $\rho_\phi \sim a^{-3}$, while amplifying fluctuations whose energy density eventually becomes comparable, $\rho_{\delta\phi} \sim \rho_{\phi}$.
The ALP quanta are dominantly produced with momenta below the ALP mass $k/a \lesssim m_\phi(\theta_i) \sim m_\phi$~(cf.~eq.~\eqref{eq:peak_momentum}), hence are non-relativistic from the moment of production.
Therefore, we can estimate the total relic ALP abundance as
\begin{equation}
    \rho_{\phi,0} = \rho_{\phi}(\Tosc) \left(\frac{\aosc}{a_0}\right)^3 \, ,
\end{equation}
where $\Tosc$ is given by eq.~\eqref{eq:Tosc} and $a_0$ is the scale factor today.
This can be expressed in terms of the critical energy density as
\begin{equation}\label{eq:Omega_ALP_general}
    h^2 \Omega_{\phi,0} = \Omega_\phi(\Tosc) \left(\frac{\aosc}{a_0}\right)^3 \left(\frac{\Hosc}{H_{100}}\right)^2\, ,
\end{equation}
with $H_{100} = 100\,\mathrm{km}/(\mathrm{Mpc}\,\mathrm{s})$.
Employing the Hubble parameter at the onset of rolling~\eqref{eq:Hosc}, we have
\begin{equation}\label{eq:Omega_phi_osc}
    \Omega_\phi(\Tosc) = \left(\frac{f_\phi^2(1-\cos(\theta_i))}{3 \Mpl^2 }\right) \left(\frac{\Tmisosc}{\Tosc}\right)^4 \left(\frac{g_\rmi{mis}}{\gosc}\right) \, .
\end{equation}
Hence, a finite trapping period enhances the relative ALP abundance, while $\Tosc = \Tmisosc$ recovers the standard expression for ALP misalignment.

Expressing the scale factor ratio in eq.~\eqref{eq:Omega_ALP_general} by temperature we find
\begin{equation}
    h^2\OmegaPhiToday = \frac{f_\phi^2}{3\Mpl^2} (1-\cos(\theta_i)) \left(\frac{T_0}{T_\rmi{osc}}\right)^3 \left(\frac{m_\phi}{H_{100}}\right)^2 \frac{g_0}{\gosc} \, ,
\end{equation}
where we have used that $\Hosc = \mphi (\Tosc/\Tmisosc)^2$.
Lower $\Tosc$, i.e., longer trapping, induces a larger present-day ALP abundance.
Via eq.~\eqref{eq:Tosc}, this can be rewritten as follows:
\begin{widetext}
\begin{equation}\label{eq:axion_DM_relic}
    h^2\Omega_{\phi,0} \simeq \begin{cases}\displaystyle
        4\times 10^{-2} \,g_\star^{-1} C_1(\delta) 
    \left(\frac{\LambdaPQ}{\mathrm{GeV}}\right)^9 \left(\frac{\mathrm{eV}}{m_\phi}\right)^4
    \left(\frac{10^{12}\,\mathrm{GeV}}{f_\phi}\right)^4 \, ,\qquad &q = 1\, , \\[1em]
    \displaystyle
    5\times 10^6 \,g_\star^{-1}  C_2 (\delta) \left(\frac{\LambdaPQ}{\mathrm{GeV}}\right)^3 \left(\frac{\mathrm{eV}}{m_\phi}\right)
    \left(\frac{10^{12}\,\mathrm{GeV}}{f_\phi}\right) \, , \qquad &q = 2\, .
    \end{cases}
\end{equation}    
\end{widetext}
Here, we have specialized to the cases $q \in \{1,2\}$. The dependence on the initial misalignment angle, or equivalently, on $\delta$, is captured by
\begin{align}
    C_1(\delta) &= (1-\cos \theta_i) \left(\frac{\cos(2\theta_i + \delta)}{-\cos(\theta_i)}\right)^3 \, , \\
    C_2(\delta) &= (1-\cos \theta_i)\left(\frac{\cos(2\theta_i + \delta)}{-\cos(\theta_i)}\right)^\frac{3}{2} \, .
\end{align}
For a set of model parameters $\{\mphi,\fphi, \delta\}$, $\LambdaPQ$ can be adjusted within the limits derived in the last section to meet today's CDM density.
Note that larger $q$ generally requires a smaller $\LambdaPQ$ to achieve a given relic abundance.

By the black lines in fig.~\ref{fig:overview_parameter_space}, we indicate the parameter combinations where $h^2\OmegaPhiToday = 0.12$ is fulfilled.
The dashed~(solid) line again corresponds to $q=1$~($q=2$).
In the ordinary misalignment mechanism, the relic abundance becomes independent of $\LambdaPQ$.
Therefore, the black curves intersect with the orange lines at the mass where the correct ALP abundance is reproduced without trapping.
Moving to smaller ALP masses, the standard misalignment mechanism would underproduce CDM.
Therefore, a finite trapping period is required to enhance the ALP abundance.
As a consequence, larger ratios $\LambdaPQ/\mphi$ are necessary, and the black lines approach the overclosure bounds.

Smaller $\fphi$ generally allow for a wider range of ALP masses due to the suppression of the ALP energy density~(cf.~eq.~\eqref{eq:Omega_phi_osc}).
For, e.g., $\fphi = 10^{17}\,\mathrm{GeV}$, we cannot reproduce the required CDM abundance in the chosen mass range.
In the next section, we will see that the GW signal is suppressed for small $\fphi$, limiting the parameter space where both sizable GW amplitudes and the correct CDM abundance is achieved.
This is a well-known problem in theories featuring ALP-induced GWs~\cite{Machado:2018nqk,Ratzinger:2020oct,Madge:2021abk,Gerlach:2025fkr}.
Note, however, that the large-$\mphi$ parameter space can be opened up through additional model building, e.g., by considering a time-varying ALP mass~\cite{McAllister:2008hb,Silverstein:2008sg,Hebecker:2014eua,McAllister:2014mpa,Blumenhagen:2014gta,Marchesano:2014mla} or a period of kination~\cite{Madge:2021abk}.

\section{Gravitational waves}
The resonant production of ALP quanta generates anisotropies in the energy momentum tensor of the cosmic fluid, i.e., leading to the generation of stochastic GWs~\cite{Chatrchyan:2020pzh,Madge:2021abk}.
In the following, we first provide simple scaling estimates of the position of the spectral peak, before discussing our numerical computation and the associated observational prospects.
\subsection{Analytic estimates}
The amplitude of a SGWB can generally be estimated as~\cite{Buchmuller:2013lra,Giblin:2014gra,Machado:2018nqk,Machado:2019xuc,Madge:2021abk}
\begin{align}\label{eq:OmegaGW_est_general}
    \Omega_{\rmii{GW},\star}^\rmi{peak} \simeq \Omega_\rmi{src}^2 \left(\frac{a_\star H_\star}{k_\rmi{peak}}\right)^2 \simeq \left(\frac{\Omega_{\phi,\star}}{2}\right)^2 \left(\frac{a_\star H_\star}{k_\rmi{peak}}\right)^2 \, .
\end{align}
Here, $\Omega_\rmi{src} \simeq \Omega_{\phi,\star}/2$ denotes the energy budget of the source, which corresponds to the ALP energy density at the time of GW production.
We have included a factor $1/2$, since the initial ALP energy density is transferred to both kinetic and gradient energy, however, only the gradient contribution acts as a GW source.
The second term amounts to a suppression factor that depends on the typical scale of the produced GWs; fluctuations on larger spatial scales induce stronger anisotropies, hence enhance the GW amplitude.
Note that for fragmentation to be efficient, the growth rate has to exceed the Hubble parameter by orders of magnitude~(cf.~eq.~\eqref{eq:m_over_H_est}). 
Therefore, ALP production takes place within a fraction of a Hubble time, and we evaluate all quantities that enter eq.~\eqref{eq:OmegaGW_est_general} at the onset of oscillations.
Then, the characteristic physical momentum scale is given by the mode which experiences the fastest growth,
\begin{equation}\label{eq:peak_momentum_estimate}
    \frac{k_\rmi{peak}}{a} \simeq \frac{\mphi}{\sqrt{2}} (1 - \cos{\theta_i})^\frac{1}{2}\, .
\end{equation}
Employing eq.~\eqref{eq:Hosc}, we then express the ratio of the Hubble rate and the physical peak momentum as
\begin{equation}
    \left(\frac{\aosc \Hosc}{k_\rmi{peak}}\right)^2 =  2 (1-\cos\theta_i)\left(\frac{\Tosc}{\Tmisosc}\right)^4 \frac{\gosc}{g_\rmi{mis}} \, ,
\end{equation}
where $\Tmisosc$ is given by eq.~\eqref{eq:T_osc,aa}.
The relative ALP energy density at the time of oscillations is given by eq.~\eqref{eq:Omega_phi_osc}.
Specializing again to the cases $q\in\{1,2\}$ and evaluating eq.~\eqref{eq:OmegaGW_est_general} yields
\begin{align}\label{eq:Omega_GW_estimate_final}
    \Omega_{\rmii{GW},\star}^\rmi{peak} \simeq 
\begin{cases}\displaystyle
    65 \,\gosc^{-1} C_3\left(\frac{\LambdaPQ^{6}}{\Mpl \fphi^2 \mphi^3}\right)^2 \, ,\quad &q=1 \, ,\\[1em]
    \displaystyle
    4 \,\gosc^{-1} C_4 \left(\frac{\LambdaPQ^2}{\mphi \Mpl}\right)^2  \, ,\quad &q=2 \, ,
\end{cases} 
\end{align}
with
\begin{align}
    C_3 &= \left(\frac{\sqrt{2}(1-\cos\theta_i)^\frac{1}{2}\cos^2(2\theta_i + \delta)}{\cos^{2}(\theta_i)}\right)^2\,, \\
    C_4 &= \left(\frac{\sqrt{2}(1-\cos(\theta_i))^\frac{1}{2} \cos(2 \theta_i + \delta)}{\cos \theta_i}\right)^2\,.
\end{align}
This allows for a direct estimate of the anticipated GW amplitude given some benchmark parameters $\{\mphi,\fphi,\LambdaPQ\}$.
However, note that the model parameters are not independent of each other, but have to be adjusted in order to remain within the viable region of parameter space. 
Decreasing $\mphi$ and $\fphi$, for instance, requires a smaller $\UPQ$ breaking scale $\LambdaPQ$ for the axion to not overclose the Universe.

\begin{figure*}
    \centering
    \includegraphics[width=\linewidth]{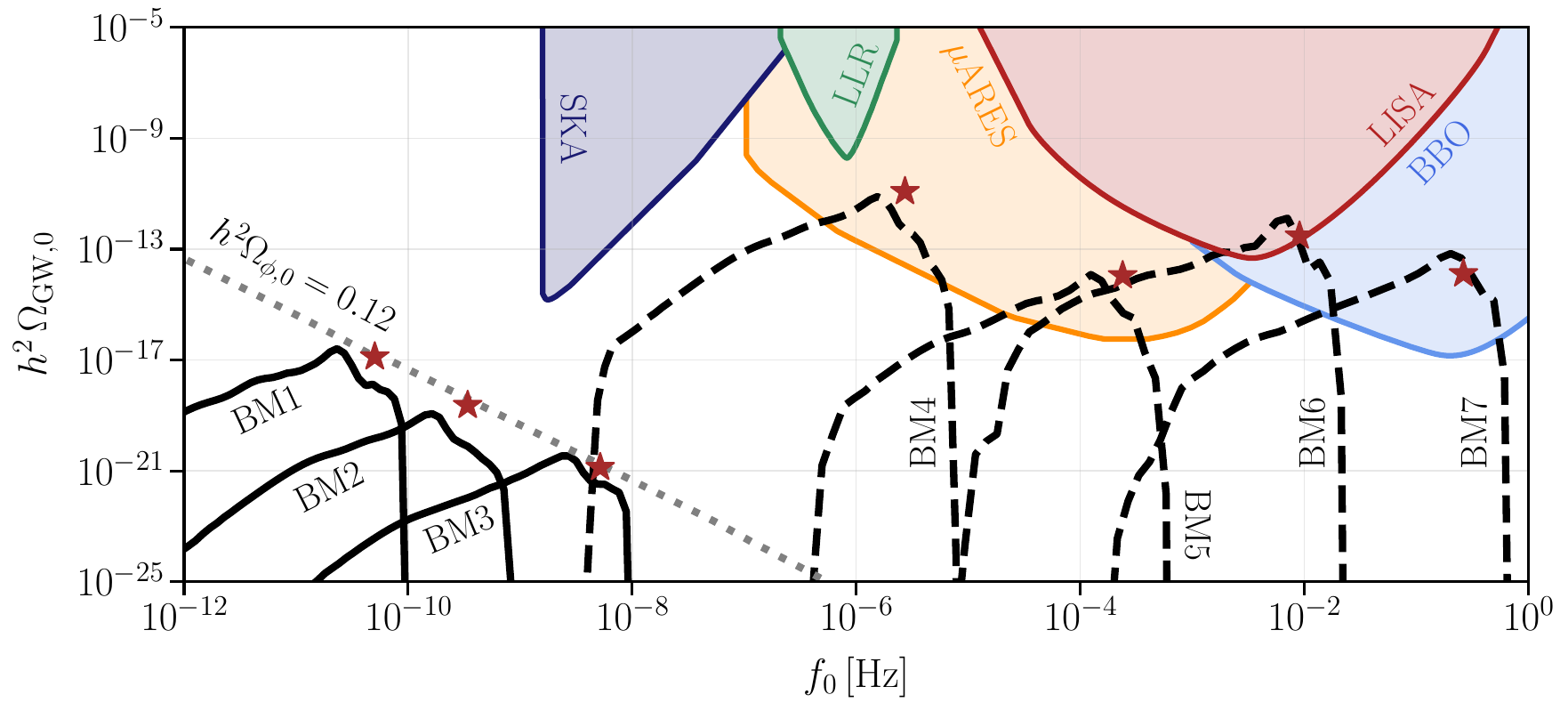}
    \caption{Projected present GW spectra for the benchmark parameters from table~\ref{tab:benchmarks}. The colored curves indicate the power-law integrated sensitivities of several future GW experiments. The black solid lines correspond to benchmarks that reproduce the correct CDM abundance. That is, ALPs are overproduced above the gray dotted line and the dashed spectra can only be realized through further model building to suppress the relic abundance. By varying the ALP mass and decay constants, the spectral peak moves into the sensitivity regions of future observatories. The red star indicates our analytic estimate of the peak~\eqref{eq:Omega_GW_estimate_final}, which correctly describes the scaling behavior of the GW signal. }
    \label{fig:observational_prospects}
\end{figure*}

\subsection{Numerical computation}
Employing our numerical analysis of eqs.~\eqref{eq:lin_eom_hom} and~\eqref{eq:lin_eom_fluc}, we can directly compute the resulting GW spectrum from the simulated mode functions~$u_k$.
The spectral GW energy density from scalar field fluctuations reads~\cite{Figueroa:2017vfa,Madge:2021abk}
\begin{equation}
    \begin{split}
    \frac{\mathrm{d}\Omega_{\rmii{GW},\star}}{\mathrm{d}\log k} &= \frac{1}{\rho_\rmi{tot}} \left(\frac{k}{4\pi^2 a_\star^2 \Mpl}\right)^2 \int\limits_0^\infty \mathrm{d}q \,q^5 \sin^4\theta \\
    &\times \int\limits_{|k-q|}^{|k+q|}\mathrm{d}l \, l \left[|I_c(\boldsymbol{k},\boldsymbol{q},\tau)|^2 |+I_s(\boldsymbol{k},\boldsymbol{q},\tau)|^2 \right]\, ,
    \end{split}
\end{equation}
where $\rho_\rmi{tot}$ denotes the total energy density in the Universe at the time of production.
Furthermore,
\begin{equation}
    l = |\boldsymbol{k} - \boldsymbol{q}| = \sqrt{k^2 - q^2 - 2kq \cos{\theta}} \, ,
\end{equation}
and the time integrals over the mode functions are given by
\begin{equation}
    I_{c/s}(\boldsymbol{k},\boldsymbol{q},\tau) = \int\limits_{\tau_\rmii{osc}}^{\tau_\star} \frac{\mathrm{d}\eta}{a(\eta)}  
    \begin{Bmatrix}
        \cos(k\eta)\\
        \sin(k\eta)
    \end{Bmatrix} u_q(\eta) u_l(\eta) \, . 
\end{equation}
For the computation we follow the strategy from~\cite{Madge:2021abk}.
That is, we discretize the integrals into sums over the simulated mode functions.
Finally, to obtain the GW spectra today, we redshift via~\cite{Kamionkowski:1993fg,Caprini:2018mtu}
\begin{align}\label{eq:redshift_GW}
    h^2 \Omega_{\rmii{GW},0} &= 1.67 \times 10^{-5} \left(\frac{100}{g_\star}\right)^\frac{1}{3} \Omega_{\rmii{GW},\star} \, , \\
    f_0 &= 1.65 \times 10^{-7} \,\mathrm{Hz}\, \frac{k}{a_\star H_\star} \frac{T_\star}{\mathrm{GeV}} \left(\frac{g_\star}{100}\right)^\frac{1}{6} \, .
\end{align}

\subsection{Results}
In fig.~\ref{fig:observational_prospects}, we show some exemplary GW spectra, computed numerically with the benchmark parameters given in table~\ref{tab:benchmarks}.
The colored areas indicate the power-law integrated sensitivity curves~\cite{Breitbach:2018ddu,Schmitz:2020syl} of the future Square Kilometre Array~(SKA)~\cite{Janssen:2014dka}, $\mu$ARES~\cite{Sesana:2019vho}, Big Bang Observer~(BBO)~\cite{Crowder:2005nr}, and LISA~\cite{2017arXiv170200786A,Robson:2018ifk,LISACosmologyWorkingGroup:2022jok}. 
In addition, we display the projected prospects for GW detection via Lunar Laser Ranging~(LLR)~\cite{Foster:2025nzf}.

The spectra plotted with solid lines correspond to benchmarks where the ALP reproduces the observed CDM abundance, $h^2\OmegaPhiToday = 0.12$.
The red stars correspond to our estimates of the peak position~\eqref{eq:Omega_GW_estimate_final}.
Note that our analytic results slightly overestimate the peak frequency and amplitude. This is because we evaluate all quantities at the onset of oscillations, i.e., neglect their redshift until the time of GW production. In addition, our simulations show that the factor $1/2$ in eq.~\eqref{eq:OmegaGW_est_general} overestimates the gradient contribution to the total fluctuation energy density.

By combining eqs.~\eqref{eq:axion_DM_relic} and \eqref{eq:Omega_GW_estimate_final}, the condition to obtain the correct relic abundance can be translated to the $f - h^2\Omega_{\rmii{GW},0}$ parameter space; this is indicated by the gray dotted line.
Above this line, ALPs are overproduced.
Hence, we do not find a parameter space that yields both detectable GW signals and ALP CDM.
This generic feature of fragmentation models was already noted in~\cite{Chatrchyan:2020pzh,Madge:2021abk} and can be understood from eq.~\eqref{eq:OmegaGW_est_general} and the scaling behavior of the ALP.
From the moment of production, the ALP quanta redshift like matter, hence are enhanced $\propto a$ compared to the background.
A larger GW frequency, i.e., an earlier onset of ALP oscillations therefore requires a smaller initial relative ALP energy density to not overclose the Universe, decreasing the GW amplitude.

In the remaining parameter space, further model building is required to suppress the relic ALP abundance. 
However, it is useful to study this region to obtain, for instance, information on the maximum GW amplitude.
Increasing the ALP mass shifts the GW spectra to higher frequencies.
The amplitude is determined by the combination of $\fphi$ and $\LambdaPQ$. 
Our peak estimate correctly captures the scaling behavior of the signal.
Interestingly, we find maximum amplitudes of $h^2 \Omega_{\rmii{GW},0} \sim \mathcal{O}(10^{-12})$, exceeding the findings from previous studies of fragmentation-induced GW signals~\cite{Chatrchyan:2020pzh,Madge:2021abk} by roughly two orders of magnitude.
This is caused by delayed onset of ALP oscillations, which increases the energy budget available to be converted into GWs.\footnote{See also~\cite{Gerlach:2025fkr} for a similar enhancement of the GW signal from axion-photon systems.}
Notably, we find a change of the spectral behavior around the peak, where $h^2 \Omega_{\rmii{GW},0} \propto k^2$, compared to ref.~\cite{Madge:2021abk}.
In addition, our spectra are peaked towards the UV, which differs from the IR-dominated signals found in~\cite{Chatrchyan:2020pzh}.
Hence, the finite-temperature contributions to the ALP potential directly affect the form of the GW signal.
However, our linearized analysis is not sufficient to make conclusive statements, as higher-order interactions between the ALP quanta alter the spectral shape.
This requires a lattice analysis, which is relegated to future work.

\begin{table}[t]
\centering
\setlength{\tabcolsep}{8pt}
\renewcommand{\arraystretch}{1.7}
\begin{tabular}{|c|c|c|c|c|c|}
    \hline
    & $\mphi\,[\mathrm{eV}]$
    & $\fphi\,[\mathrm{GeV}]$
    & $\LambdaPQ\,[\mathrm{GeV}]$
    & $\delta$
    & $q$
    \\ \hline\hline
    BM1
    & $10^{-19}$
    & $10^{14}$
    & $7.5\times 10^{-9}$
    & $1$
    & $2$
    \\ \hline
    BM2
    & $10^{-17}$
    & $5\times 10^{13}$
    & $3.1\times 10^{-8}$
    & $0.1$
    & $2$
    \\ \hline
    BM3
    & $10^{-15}$
    & $10^{13}$
    & $7.6\times 10^{-8}$
    & $1$
    & $2$
    \\ \hline
    BM4 
    & $10^{-9}$
    & $5\times 10^{15}$
    & $0.048$
    & $0.1$
    & $1$
    \\ \hline
    BM5 
    & $10^{-5}$
    & $10^{15}$
    & $1.2$
    & $0.01$
    & $2$
    \\ \hline
    BM6
    & $10^{-2}$
    & $2\times 10^{15}$
    & $100$
    & $0.3$
    & $2$
    \\ \hline
    BM7
    & $10$
    & $10^{15}$
    & $1980$
    & $0.1$
    & $1$
    \\ \hline
\end{tabular}
\caption{%
    Benchmark parameters employed for the GW spectra in fig.~\ref{fig:observational_prospects}.
    }
\label{tab:benchmarks}
\end{table}

\section{Discussion and conclusions}
\label{sec:discussion}

In this work, we have investigated the generation of a SGWB through ALP fragmentation in finite-temperature trapped misalignment models.
Below, we summarize our key findings, discuss their broader implications, and outline directions for future work.

We have shown that explicit PQ breaking through finite-temperature contributions to the potential traps the axion field in a metastable minimum. 
Once the thermal barrier disappears, the field is released and undergoes rapid oscillations around the true minimum.
The finite ALP velocity triggers a resonance in the equation of motion, leading to the amplification of long-wavelength fluctuations.
If the trapping period is sufficiently long, the resulting field fragmentation produces a GW background analogous to that from conventional axion fragmentation~\cite{Chatrchyan:2020pzh,Madge:2021abk}, along with an inhomogeneous axion field population. 
Our analytic estimates and numerical analysis show that the GW amplitude is enhanced by up to two orders of magnitude relative to the standard fragmentation case. 
This is a direct consequence of the delayed onset of oscillations; the GW amplitude grows with the length of the trapping period.
The GW peak frequency is mainly determined by the ALP mass $\mphi$.
The enhancement of the amplitude implies that the GWs are within the reach of future interferometers such as LISA, BBO, and $\mu$Ares, as well as pulsar timing arrays~(see fig.~\ref{fig:observational_prospects}).
However, this requires an additional mechanism to suppress the relic ALP abundance.

The spectral shape displays a $k^{2}$ scaling around the peak, which changes to $k^{3}$ in the IR before the rapid falloff. 
Our signal peaks in the UV, whereas previous works found an IR-dominated peak in the ALP monodromy scenario~\cite{Chatrchyan:2020pzh}. 
This reveals that the GW spectra are sensitive to the release dynamics and the shape of the finite-temperature potential.
These spectral features of the GWs may provide a discriminant between different ALP-induced GW signals.

Finally, we comment on future directions based on the findings of this work. 
Our linearized analysis describes the main qualitative features of the ALP dynamics, however, a complete nonlinear treatment is required to conclude these predictions.\footnote{Recent lattice study of $\mathcal{Z}_{N}$ QCD axions~\cite{Co:2025jnj}, which was one of the original model motivations for trapped misalignment, affirms the necessity of simulations.} 
Lattice simulations of the post-release dynamics include mode–mode couplings, which induce spectral broadening. 
Capturing these nonlinear effects along with the backreaction of amplified modes on the zero mode provides a quantitative prediction of the GW spectrum. 

Considering the QCD axion, where temperature-dependent instanton effects may induce similar trapping dynamics, is a natural extension.
Furthermore, one may adopt other trapping potentials arising from, for instance, interactions with a dark pure Yang-Mills sector, employing the free energy calculated in~\cite{Arnold:1994ps}. 
These types of dark sectors are often realized in string theory-inspired axion models within the axiverse.  
Such axion models have also been proposed as a framework for realizing high-quality QCD axions; see, for instance,~\cite{ZambujalFerreira:2021cte}.

Lastly, since most of the parameter space is constrained by the CDM abundance in the minimal setup, it would be interesting to explore potential mechanisms to alleviate the overproduction of ALPs.
One example was suggested in~\cite{Chatrchyan:2020pzh}, which considers an additional relativistic phase after ALP fragmentation to dilute the relic ALP abundance.

In summary, the resonant behavior emerging from thermally trapped misalignment represents a robust mechanism for generating SGWBs in the early Universe. The distinctive spectral features and amplitude enhancements found here motivate future theoretical, numerical, and observational efforts to probe the rich dynamics of axion-like sectors through their gravitational-wave signatures.

\textbf{Acknowledgements.} We thank Dieter~B. in Mainz that initially sparked this work.
We further thank C.~Gerlach, E.~Morgante, W.~Ratzinger, and P.~Schwaller for comments and discussion of the manuscript.
NR thanks P.~Sørensen for illuminating him into thinking about trapped axions. NR acknowledges support in part by the European Union - NextGenerationEU through the PRIN Project ``Charting unexplored avenues in Dark Matter'' (20224JR28W), and INFN TAsP. 

\section*{\Appendix}\label{sec:appendix}
In this section, we discuss our numerical approach to solving the ALP equations of motion.
\subsection{Solving the equations of motion}
\begin{figure}
    \centering
    \includegraphics[width=\linewidth]{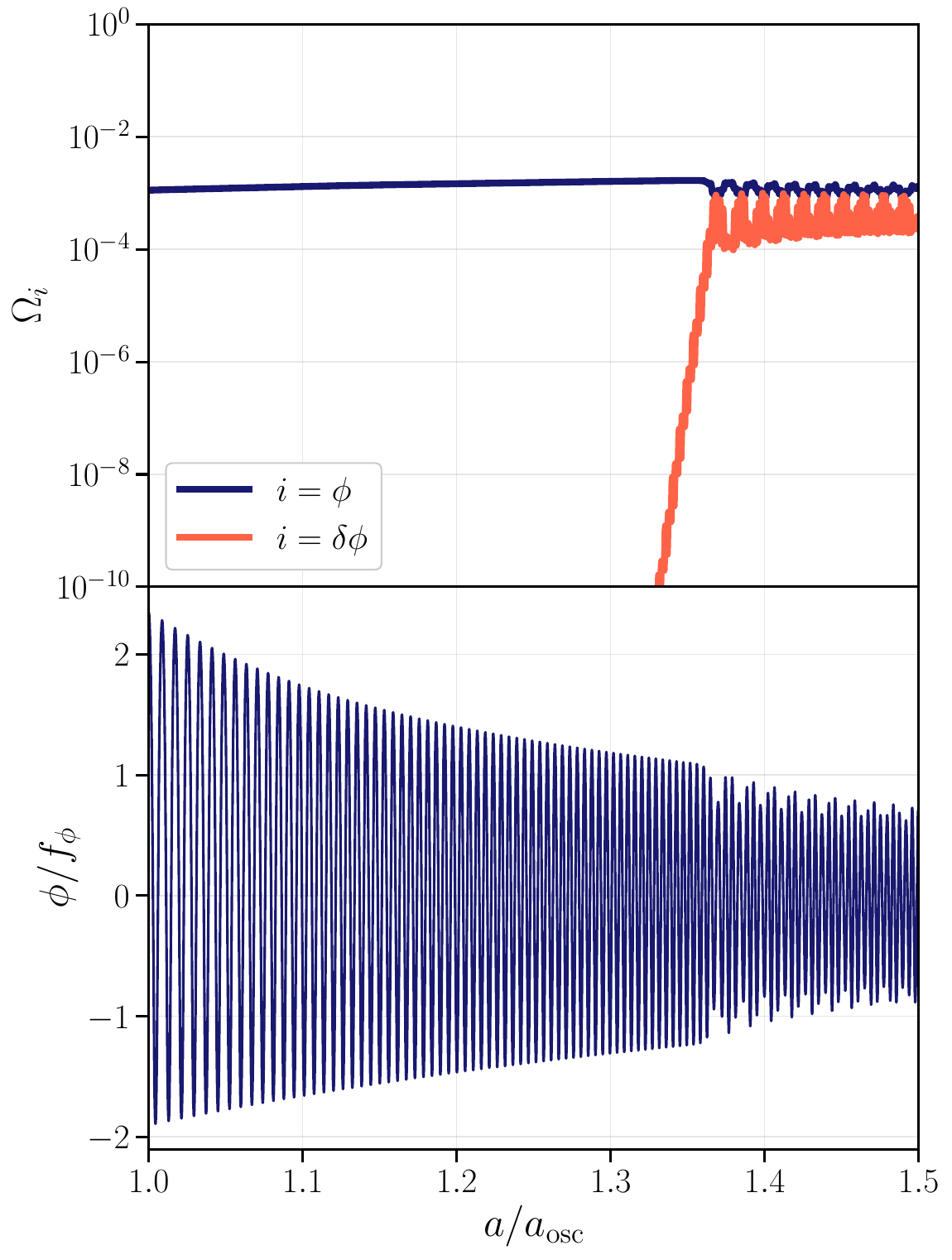}
    \caption{Exemplary simulation of the fragmentation process, employing $m_\phi = 10^{-19}\,\mathrm{eV}$, $f_\phi = $, $\LambdaPQ \approx 7.5\times 10^{-9}\,\mathrm{GeV}$, $\delta = 1$, $n=2$, and $q=2$. Top: Energy densities of the zero mode~(blue) and the fluctuations~(orange), normalized to the total energy density of the Universe, as a function of the scale factor. Bottom: Oscillation amplitude of the zero mode. Around $a \approx 1.36 \aosc$, the fluctuation energy density becomes comparable to the one of the zero mode. Then, backreaction sets in and dampens the oscillations.}
    \label{fig:benchmark_sim}
\end{figure}

Given a set of input parameters $\{\mphi,\fphi,\LambdaPQ,\delta,n,q\}$, we solve the coupled eqs.~\eqref{eq:lin_eom_hom} and~\eqref{eq:lin_eom_fluc} numerically for $N_k = 10^4$ ALP modes.
The result for a benchmark that reproduces the correct CDM abundance is shown in fig.~\ref{fig:benchmark_sim}.
Here, the upper panel displays the relative energy density of the zero mode and the fluctuations, while the second panel shows the oscillation amplitude.
Initially, the energy density of the zero mode is merely affected by the cosmic expansion, i.e. $\Omega_\phi \propto a$.
During each oscillation cycle, unstable modes receive an amplification, increasing the fluctuation energy density which eventually becomes comparable to that of the zero mode, $\Omega_{\delta \varphi} \sim \Omega_\phi$. 
For the given benchmark this occurs at $a \approx 1.37 \aosc$.
Subsequently, backreaction from the produced quanta on the zero mode becomes important, governed by the third derivative of the potential in eq.~\eqref{eq:lin_eom_hom}.
We observe that this leads to an oscillating behavior of the energy densities~(cf.~upper panel).
As the zero-mode energy density decreases upon the impact of the fluctuations, the oscillation amplitude decreases~(cf.~lower panel).
This slows down the zero mode; i.e., the instability band becomes narrower and eventually vanishes. 

\begin{figure}
    \centering
    \includegraphics[width=\linewidth]{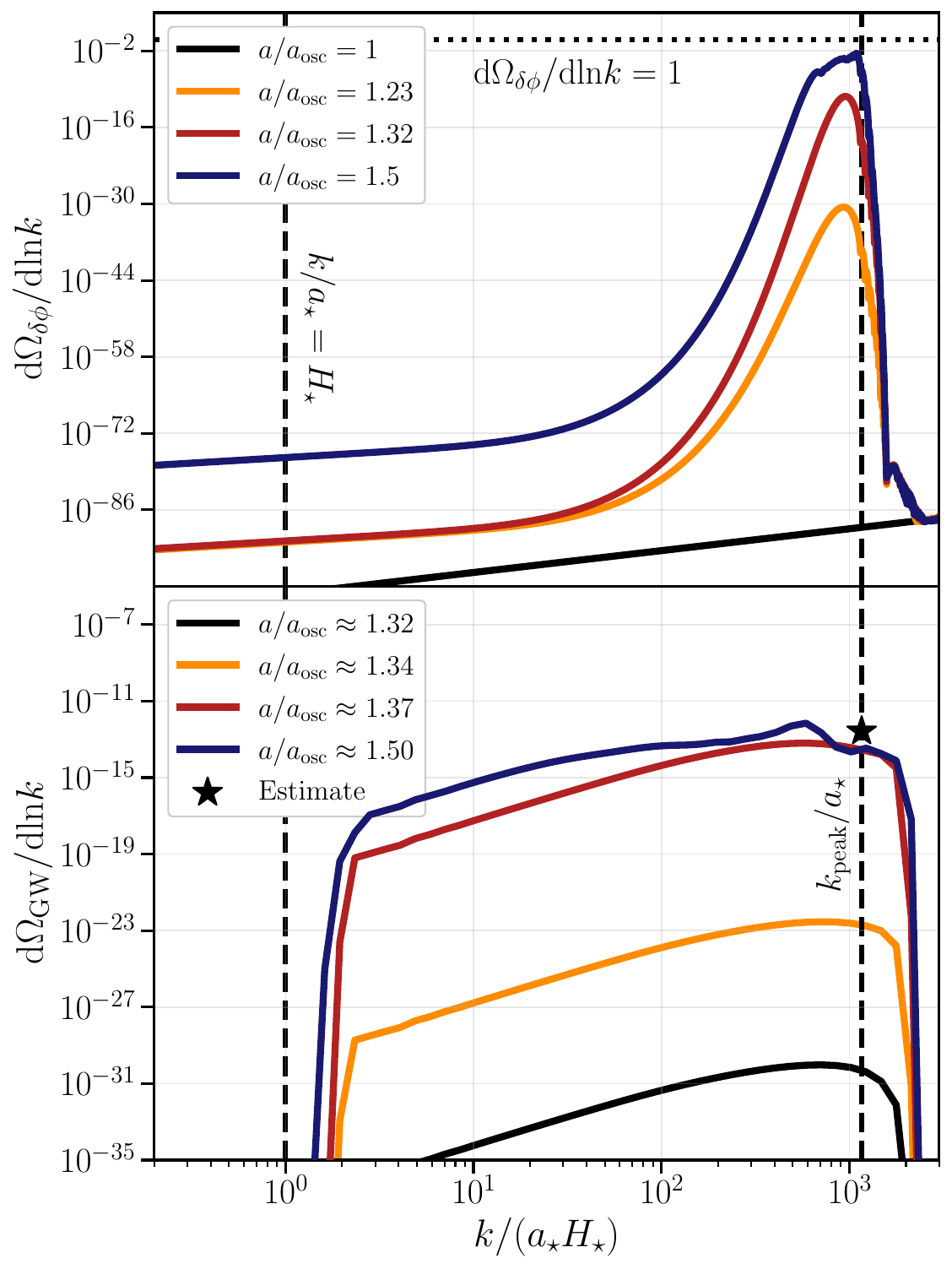}
    \caption{Spectral ALP~(top) and GW~(bottom) energy density at different times during the simulation, using the model parameters from fig.~\ref{fig:benchmark_sim}. We display the spectra as a function of the physical momenta $k/a_\star$, normalized to the Hubble rate at the time of production. The dashed line in the UV corresponds to our estimate of the fastest growing mode~\eqref{eq:peak_momentum}, which also sets the peak of the GW spectrum.}
    \label{fig:benchmark_spectra}
\end{figure}

Note that once the fluctuations grow large, our perturbative approach necessarily induces uncertainties as the expansion~\eqref{eq:potential_expansion} breaks down.
Higher-order effects that are not captured by our simulation, such as interactions between the produces ALP quanta, become important.
To this end, classical lattice simulations are required~\cite{Chatrchyan:2020pzh}, which is beyond the scope of our work.

Fig.~\ref{fig:benchmark_spectra} shows the associated energy spectra of the fluctuations and GWs at different times during the simulation, as a function of the physical momentum normalized to the Hubble parameter at the start of the simulation.
The GW spectrum is computed numerically from the extracted mode functions, as described in the main body.
GWs are predominantly produced during the initial stages of the evolution, as shown by the red benchmark in the lower panel.
This benchmark approximately corresponds to the moment where backreaction sets in, which, however, has a minor impact on the GW spectrum.
On the lattice, where the full backreaction is incorporated, the GW spectrum typically broadens due to interactions among the modes.
As we are mostly interested in estimating the position of the resulting GW peak, we consider our results robust.
Finally, let us point out the remarkable agreement between the numerical computation and the analytical estimate of the GW peak frequency~(eq.~\eqref{eq:peak_momentum_estimate}), and amplitude~(eq.~\eqref{eq:Omega_GW_estimate_final}), indicated by the orange star in the lower panel.

%%%%%%%%%%%%%%%%%%%%%%%% Bibliography %%%%%
\bibliographystyle{apsrev4-1}
\bibliography{biblio.bib}

@article{DiLuzio:2024fyt,
    author = "Di Luzio, Luca and S{\o}rensen, Philip",
    title = "{Axion production via trapped misalignment from Peccei-Quinn symmetry breaking}",
    eprint = "2408.04623",
    archivePrefix = "arXiv",
    primaryClass = "hep-ph",
    doi = "10.1007/JHEP10(2024)239",
    journal = "JHEP",
    volume = "10",
    pages = "239",
    year = "2024"
}

@article{DiLuzio:2021pxd,
    author = "Di Luzio, Luca and Gavela, Belen and Quilez, Pablo and Ringwald, Andreas",
    title = "{An even lighter QCD axion}",
    eprint = "2102.00012",
    archivePrefix = "arXiv",
    primaryClass = "hep-ph",
    reportNumber = "DESY-21-010, DESY 21-010, IFT-UAM/CSIC-20-143, FTUAM-20-21",
    doi = "10.1007/JHEP05(2021)184",
    journal = "JHEP",
    volume = "05",
    pages = "184",
    year = "2021"
}

@article{Schmitt:2024pby,
    author = "Schmitt, Daniel and Sagunski, Laura",
    title = "{QCD-sourced tachyonic phase transition in a supercooled Universe}",
    eprint = "2409.05851",
    archivePrefix = "arXiv",
    primaryClass = "hep-ph",
    doi = "10.1088/1475-7516/2025/02/075",
    journal = "JCAP",
    volume = "02",
    pages = "075",
    year = "2025"
}

@article{Dutka:2025oqt,
    author = "Dutka, Tomasz P. and Jung, Tae Hyun and Shin, Chang Sub",
    title = "{What happens when supercooling is terminated by curvature flipping of the effective potential?}",
    eprint = "2412.15864",
    archivePrefix = "arXiv",
    primaryClass = "hep-ph",
    reportNumber = "CTPU-PTC-24-40",
    doi = "10.1007/JHEP05(2025)182",
    journal = "JHEP",
    volume = "05",
    pages = "182",
    year = "2025"
}

@article{Hook:2018jle,
    author = "Hook, Anson",
    title = "{Solving the Hierarchy Problem Discretely}",
    eprint = "1802.10093",
    archivePrefix = "arXiv",
    primaryClass = "hep-ph",
    doi = "10.1103/PhysRevLett.120.261802",
    journal = "Phys. Rev. Lett.",
    volume = "120",
    number = "26",
    pages = "261802",
    year = "2018"
}

@article{Svrcek:2006yi,
    author = "Svrcek, Peter and Witten, Edward",
    title = "{Axions In String Theory}",
    eprint = "hep-th/0605206",
    archivePrefix = "arXiv",
    reportNumber = "SLAC-PUB-11894",
    doi = "10.1088/1126-6708/2006/06/051",
    journal = "JHEP",
    volume = "06",
    pages = "051",
    year = "2006"
}

@article{Arvanitaki:2009fg,
    author = "Arvanitaki, Asimina and Dimopoulos, Savas and Dubovsky, Sergei and Kaloper, Nemanja and March-Russell, John",
    title = "{String Axiverse}",
    eprint = "0905.4720",
    archivePrefix = "arXiv",
    primaryClass = "hep-th",
    doi = "10.1103/PhysRevD.81.123530",
    journal = "Phys. Rev. D",
    volume = "81",
    pages = "123530",
    year = "2010"
}

@article{Marsh:2015xka,
    author = "Marsh, David J. E.",
    title = "{Axion Cosmology}",
    eprint = "1510.07633",
    archivePrefix = "arXiv",
    primaryClass = "astro-ph.CO",
    reportNumber = "KCL-PH-TH-2015-50",
    doi = "10.1016/j.physrep.2016.06.005",
    journal = "Phys. Rept.",
    volume = "643",
    pages = "1--79",
    year = "2016"
}

@article{Arnold:1994ps,
    author = "Arnold, Peter Brockway and Zhai, Cheng-Xing",
    title = "{The Three loop free energy for pure gauge QCD}",
    eprint = "hep-ph/9408276",
    archivePrefix = "arXiv",
    reportNumber = "UW-PT-94-03",
    doi = "10.1103/PhysRevD.50.7603",
    journal = "Phys. Rev. D",
    volume = "50",
    pages = "7603--7623",
    year = "1994"
}

@article{ZambujalFerreira:2021cte,
    author = "Zambujal Ferreira, Ricardo and Notari, Alessio and Pujol{\`a}s, Oriol and Rompineve, Fabrizio",
    title = "{High Quality QCD Axion at Gravitational Wave Observatories}",
    eprint = "2107.07542",
    archivePrefix = "arXiv",
    primaryClass = "hep-ph",
    doi = "10.1103/PhysRevLett.128.141101",
    journal = "Phys. Rev. Lett.",
    volume = "128",
    number = "14",
    pages = "141101",
    year = "2022"
}

@article{Benabou:2024msj,
    author = "Benabou, Joshua N. and Buschmann, Malte and Foster, Joshua W. and Safdi, Benjamin R.",
    title = "{Axion Mass Prediction from Adaptive Mesh Refinement Cosmological Lattice Simulations}",
    eprint = "2412.08699",
    archivePrefix = "arXiv",
    primaryClass = "hep-ph",
    reportNumber = "FERMILAB-PUB-24-0912-T",
    doi = "10.1103/6v21-d6sj",
    journal = "Phys. Rev. Lett.",
    volume = "134",
    number = "24",
    pages = "241003",
    year = "2025"
}

@article{Gorghetto:2020qws,
    author = "Gorghetto, Marco and Hardy, Edward and Villadoro, Giovanni",
    title = "{More axions from strings}",
    eprint = "2007.04990",
    archivePrefix = "arXiv",
    primaryClass = "hep-ph",
    doi = "10.21468/SciPostPhys.10.2.050",
    journal = "SciPost Phys.",
    volume = "10",
    number = "2",
    pages = "050",
    year = "2021"
}

@article{Correia:2024cpk,
    author = "Correia, Jos{\'e} and Hindmarsh, Mark and Lizarraga, Joanes and Lopez-Eiguren, Asier and Rummukainen, Kari and Urrestilla, Jon",
    title = "{Scaling density of axion strings in terasite simulations}",
    eprint = "2410.18064",
    archivePrefix = "arXiv",
    primaryClass = "hep-ph",
    doi = "10.1103/PhysRevD.111.063532",
    journal = "Phys. Rev. D",
    volume = "111",
    number = "6",
    pages = "063532",
    year = "2025"
}

@article{Saikawa:2024bta,
    author = "Saikawa, Ken'ichi and Redondo, Javier and Vaquero, Alejandro and Kaltschmidt, Mathieu",
    title = "{Spectrum of global string networks and the axion dark matter mass}",
    eprint = "2401.17253",
    archivePrefix = "arXiv",
    primaryClass = "hep-ph",
    reportNumber = "KANAZAWA-24-02, MPP-2024-18",
    doi = "10.1088/1475-7516/2024/10/043",
    journal = "JCAP",
    volume = "10",
    pages = "043",
    year = "2024"
}

@article{Co:2025jnj,
    author = "Co, Raymond T. and Lee, Taegyu and Leonard, Owen P.",
    title = "{(Non-)Perturbative Dynamics of a Light QCD Axion: Dark Matter and the Strong CP Problem}",
    eprint = "2508.00979",
    archivePrefix = "arXiv",
    primaryClass = "hep-ph",
    reportNumber = "CETUP2025-00",
    month = "8",
    year = "2025"
}

@article{Kim:1986ax,
    author = "Kim, Jihn E.",
    title = "{Light Pseudoscalars, Particle Physics and Cosmology}",
    reportNumber = "SNUHE-86-09",
    doi = "10.1016/0370-1573(87)90017-2",
    journal = "Phys. Rept.",
    volume = "150",
    pages = "1--177",
    year = "1987"
}

@article{DiLuzio:2021gos,
    author = "Di Luzio, Luca and Gavela, Belen and Quilez, Pablo and Ringwald, Andreas",
    title = "{Dark matter from an even lighter QCD axion: trapped misalignment}",
    eprint = "2102.01082",
    archivePrefix = "arXiv",
    primaryClass = "hep-ph",
    reportNumber = "DESY 21-011, DESY-21-011, IFT-UAM/CSIC-20-144, FTUAM-20-21",
    doi = "10.1088/1475-7516/2021/10/001",
    journal = "JCAP",
    volume = "10",
    pages = "001",
    year = "2021"
}

@article{Gerlach:2025fkr,
    author = "Gerlach, Christopher and Schmitt, Daniel and Schwaller, Pedro",
    title = "{Supercooled Audible Axions}",
    eprint = "2504.05386",
    archivePrefix = "arXiv",
    primaryClass = "hep-ph",
    reportNumber = "MITP-25-028",
    month = "4",
    year = "2025"
}

@article{Fonseca:2019ypl,
    author = "Fonseca, Nayara and Morgante, Enrico and Sato, Ryosuke and Servant, G{\'e}raldine",
    title = "{Axion fragmentation}",
    eprint = "1911.08472",
    archivePrefix = "arXiv",
    primaryClass = "hep-ph",
    reportNumber = "DESY 19-202, DESY-19-202, MITP/19-079",
    doi = "10.1007/JHEP04(2020)010",
    journal = "JHEP",
    volume = "04",
    pages = "010",
    year = "2020"
}

@article{Madge:2021abk,
    author = "Madge, Eric and Ratzinger, Wolfram and Schmitt, Daniel and Schwaller, Pedro",
    title = "{Audible axions with a booster: Stochastic gravitational waves from rotating ALPs}",
    eprint = "2111.12730",
    archivePrefix = "arXiv",
    primaryClass = "hep-ph",
    reportNumber = "MITP-21-063",
    doi = "10.21468/SciPostPhys.12.5.171",
    journal = "SciPost Phys.",
    volume = "12",
    number = "5",
    pages = "171",
    year = "2022"
}

@article{Peccei:1977hh,
  author         = "Peccei, Roberto D. and Quinn, Helen R.",
  title          = "{CP Conservation in the Presence of Instantons}",
  journal        = "Phys. Rev. Lett.",
  volume         = "38",
  year           = "1977",
  pages          = "1440--1443",
  doi            = "10.1103/PhysRevLett.38.1440"
}

@article{Chatrchyan:2020pzh,
    author = "Chatrchyan, Aleksandr and Jaeckel, Joerg",
    title = "{Gravitational waves from the fragmentation of axion-like particle dark matter}",
    eprint = "2004.07844",
    archivePrefix = "arXiv",
    primaryClass = "hep-ph",
    doi = "10.1088/1475-7516/2021/02/003",
    journal = "JCAP",
    volume = "02",
    pages = "003",
    year = "2021"
}

@article{Weinberg:1977ma,
  author         = "Weinberg, Steven",
  title          = "{A New Light Boson?}",
  journal        = "Phys. Rev. Lett.",
  volume         = "40",
  year           = "1978",
  pages          = "223--226",
  doi            = "10.1103/PhysRevLett.40.223"
}

@article{Wilczek:1977pj,
  author         = "Wilczek, Frank",
  title          = "{Problem of Strong $P$ and $T$ Invariance in the Presence of Instantons}",
  journal        = "Phys. Rev. Lett.",
  volume         = "40",
  year           = "1978",
  pages          = "279--282",
  doi            = "10.1103/PhysRevLett.40.279"
}

@article{Preskill:1982cy,
  author         = "Preskill, John and Wise, Mark B. and Wilczek, Frank",
  title          = "{Cosmology of the Invisible Axion}",
  journal        = "Phys. Lett. B",
  volume         = "120",
  year           = "1983",
  pages          = "127--132",
  doi            = "10.1016/0370-2693(83)90637-8"
}

@article{Abbott:1982af,
  author         = "Abbott, L. F. and Sikivie, P.",
  title          = "{A Cosmological Bound on the Invisible Axion}",
  journal        = "Phys. Lett. B",
  volume         = "120",
  year           = "1983",
  pages          = "133--136",
  doi            = "10.1016/0370-2693(83)90638-X"
}

@article{Dine:1982ah,
  author         = "Dine, Michael and Fischler, Willy",
  title          = "{The Not So Harmless Axion}",
  journal        = "Phys. Lett. B",
  volume         = "120",
  year           = "1983",
  pages          = "137--141",
  doi            = "10.1016/0370-2693(83)90639-1"
}

@article{Planck:2018vyg,
    author = "Aghanim, N. and others",
    collaboration = "Planck",
    title = "{Planck 2018 results. VI. Cosmological parameters}",
    eprint = "1807.06209",
    archivePrefix = "arXiv",
    primaryClass = "astro-ph.CO",
    doi = "10.1051/0004-6361/201833910",
    journal = "Astron. Astrophys.",
    volume = "641",
    pages = "A6",
    year = "2020",
    note = "[Erratum: Astron.Astrophys. 652, C4 (2021)]"
}

@article{Machado:2018nqk,
    author = "Machado, Camila S. and Ratzinger, Wolfram and Schwaller, Pedro and Stefanek, Ben A.",
    title = "{Audible Axions}",
    eprint = "1811.01950",
    archivePrefix = "arXiv",
    primaryClass = "hep-ph",
    reportNumber = "MITP/18-107",
    doi = "10.1007/JHEP01(2019)053",
    journal = "JHEP",
    volume = "01",
    pages = "053",
    year = "2019"
}

@article{Ratzinger:2020oct,
    author = "Ratzinger, Wolfram and Schwaller, Pedro and Stefanek, Ben A.",
    title = "{Gravitational Waves from an Axion-Dark Photon System: A Lattice Study}",
    eprint = "2012.11584",
    archivePrefix = "arXiv",
    primaryClass = "astro-ph.CO",
    reportNumber = "MITP-20-086, ZU-TH-57/20",
    doi = "10.21468/SciPostPhys.11.1.001",
    journal = "SciPost Phys.",
    volume = "11",
    pages = "001",
    year = "2021"
}

@article{Figueroa:2017vfa,
    author = "Figueroa, Daniel G. and Torrenti, Francisco",
    title = "{Gravitational wave production from preheating: parameter dependence}",
    eprint = "1707.04533",
    archivePrefix = "arXiv",
    primaryClass = "astro-ph.CO",
    reportNumber = "CERN-TH-2017-152, IFT-UAM-CSIC-17-069",
    doi = "10.1088/1475-7516/2017/10/057",
    journal = "JCAP",
    volume = "10",
    pages = "057",
    year = "2017"
}

@article{Caprini:2018mtu,
    author = "Caprini, Chiara and Figueroa, Daniel G.",
    title = "{Cosmological Backgrounds of Gravitational Waves}",
    eprint = "1801.04268",
    archivePrefix = "arXiv",
    primaryClass = "astro-ph.CO",
    doi = "10.1088/1361-6382/aac608",
    journal = "Class. Quant. Grav.",
    volume = "35",
    number = "16",
    pages = "163001",
    year = "2018"
}

@article{Buchmuller:2013lra,
    author = {Buchm\"uller, Wilfried and Domcke, Valerie and Kamada, Kohei and Schmitz, Kai},
    title = "{The Gravitational Wave Spectrum from Cosmological $B-L$ Breaking}",
    eprint = "1305.3392",
    archivePrefix = "arXiv",
    primaryClass = "hep-ph",
    reportNumber = "DESY-13-050, IPMU-13-0091",
    doi = "10.1088/1475-7516/2013/10/003",
    journal = "JCAP",
    volume = "10",
    pages = "003",
    year = "2013"
}

@article{Giblin:2014gra,
    author = "Giblin, John T. and Thrane, Eric",
    title = "{Estimates of maximum energy density of cosmological gravitational-wave backgrounds}",
    eprint = "1410.4779",
    archivePrefix = "arXiv",
    primaryClass = "gr-qc",
    doi = "10.1103/PhysRevD.90.107502",
    journal = "Phys. Rev. D",
    volume = "90",
    number = "10",
    pages = "107502",
    year = "2014"
}

@article{Machado:2019xuc,
    author = "Machado, Camila S. and Ratzinger, Wolfram and Schwaller, Pedro and Stefanek, Ben A.",
    title = "{Gravitational wave probes of axionlike particles}",
    eprint = "1912.01007",
    archivePrefix = "arXiv",
    primaryClass = "hep-ph",
    reportNumber = "MITP/19-083",
    doi = "10.1103/PhysRevD.102.075033",
    journal = "Phys. Rev. D",
    volume = "102",
    number = "7",
    pages = "075033",
    year = "2020"
}

@article{Kamionkowski:1993fg,
    author = "Kamionkowski, Marc and Kosowsky, Arthur and Turner, Michael S.",
    title = "{Gravitational radiation from first order phase transitions}",
    eprint = "astro-ph/9310044",
    archivePrefix = "arXiv",
    reportNumber = "IASSNS-HEP-93-44, FERMILAB-PUB-93-235-A",
    doi = "10.1103/PhysRevD.49.2837",
    journal = "Phys. Rev. D",
    volume = "49",
    pages = "2837--2851",
    year = "1994"
}

@article{Kofman:1997yn,
    author = "Kofman, Lev and Linde, Andrei D. and Starobinsky, Alexei A.",
    title = "{Towards the theory of reheating after inflation}",
    eprint = "hep-ph/9704452",
    archivePrefix = "arXiv",
    reportNumber = "IFA-97-28, SU-ITP-97-18",
    doi = "10.1103/PhysRevD.56.3258",
    journal = "Phys. Rev. D",
    volume = "56",
    pages = "3258--3295",
    year = "1997"
}

@article{Breitbach:2018ddu,
    author = "Breitbach, Moritz and Kopp, Joachim and Madge, Eric and Opferkuch, Toby and Schwaller, Pedro",
    title = "{Dark, Cold, and Noisy: Constraining Secluded Hidden Sectors with Gravitational Waves}",
    eprint = "1811.11175",
    archivePrefix = "arXiv",
    primaryClass = "hep-ph",
    reportNumber = "CERN-TH-2018-255, MITP/18-115",
    doi = "10.1088/1475-7516/2019/07/007",
    journal = "JCAP",
    volume = "07",
    pages = "007",
    year = "2019"
}

@article{Schmitz:2020syl,
    author = "Schmitz, Kai",
    title = "{New Sensitivity Curves for Gravitational-Wave Signals from Cosmological Phase Transitions}",
    eprint = "2002.04615",
    archivePrefix = "arXiv",
    primaryClass = "hep-ph",
    reportNumber = "CERN-TH-2020-018",
    doi = "10.1007/JHEP01(2021)097",
    journal = "JHEP",
    volume = "01",
    pages = "097",
    year = "2021"
}

@article{Foster:2025nzf,
    author = "Foster, Joshua W. and Blas, Diego and Bourgoin, Adrien and Hees, Aurelien and Herrero-Valea, M{\'\i}riam and Jenkins, Alexander C. and Xue, Xiao",
    title = "{Discovering $\mu$Hz gravitational waves and ultra-light dark matter with binary resonances}",
    eprint = "2504.15334",
    archivePrefix = "arXiv",
    primaryClass = "astro-ph.CO",
    reportNumber = "FERMILAB-PUB-25-0091-T",
    month = "4",
    year = "2025"
}

@article{Janssen:2014dka,
    author = "Janssen, Gemma and others",
    editor = "Bourke, Tyler L. and others",
    title = "{Gravitational wave astronomy with the SKA}",
    eprint = "1501.00127",
    archivePrefix = "arXiv",
    primaryClass = "astro-ph.IM",
    doi = "10.22323/1.215.0037",
    journal = "PoS",
    volume = "AASKA14",
    pages = "037",
    year = "2015"
}

@article{Robson:2018ifk,
    author = "Robson, Travis and Cornish, Neil J. and Liu, Chang",
    title = "{The construction and use of LISA sensitivity curves}",
    eprint = "1803.01944",
    archivePrefix = "arXiv",
    primaryClass = "astro-ph.HE",
    doi = "10.1088/1361-6382/ab1101",
    journal = "Class. Quant. Grav.",
    volume = "36",
    number = "10",
    pages = "105011",
    year = "2019"
}

@ARTICLE{2017arXiv170200786A,
       author = {{LISA collaboration}},
        title = "{Laser Interferometer Space Antenna}",
      journal = {arXiv e-prints},
         year = 2017,
        month = feb,
          eid = {arXiv:1702.00786},
        pages = {arXiv:1702.00786},
archivePrefix = {arXiv},
       eprint = {1702.00786},
 primaryClass = {astro-ph.IM},
       adsurl = {https://ui.adsabs.harvard.edu/abs/2017arXiv170200786A}
}

@article{Crowder:2005nr,
    author = "Crowder, Jeff and Cornish, Neil J.",
    title = "{Beyond LISA: Exploring future gravitational wave missions}",
    eprint = "gr-qc/0506015",
    archivePrefix = "arXiv",
    doi = "10.1103/PhysRevD.72.083005",
    journal = "Phys. Rev. D",
    volume = "72",
    pages = "083005",
    year = "2005"
}

@article{Sesana:2019vho,
    author = "Sesana, Alberto and others",
    title = "{Unveiling the gravitational universe at $\mu$-Hz frequencies}",
    eprint = "1908.11391",
    archivePrefix = "arXiv",
    primaryClass = "astro-ph.IM",
    doi = "10.1007/s10686-021-09709-9",
    journal = "Exper. Astron.",
    volume = "51",
    number = "3",
    pages = "1333--1383",
    year = "2021"
}

@article{LISACosmologyWorkingGroup:2022jok,
    author = "Auclair, Pierre and others",
    collaboration = "LISA Cosmology Working Group",
    title = "{Cosmology with the Laser Interferometer Space Antenna}",
    eprint = "2204.05434",
    archivePrefix = "arXiv",
    primaryClass = "astro-ph.CO",
    reportNumber = "LISA CosWG-22-03, FERMILAB-PUB-22-349-SCD",
    doi = "10.1007/s41114-023-00045-2",
    journal = "Living Rev. Rel.",
    volume = "26",
    number = "1",
    pages = "5",
    year = "2023"
}

@article{Dufaux:2007pt,
    author = "Dufaux, Jean Francois and Bergman, Amanda and Felder, Gary N. and Kofman, Lev and Uzan, Jean-Philippe",
    title = "{Theory and Numerics of Gravitational Waves from Preheating after Inflation}",
    eprint = "0707.0875",
    archivePrefix = "arXiv",
    primaryClass = "astro-ph",
    doi = "10.1103/PhysRevD.76.123517",
    journal = "Phys. Rev. D",
    volume = "76",
    pages = "123517",
    year = "2007"
}

@article{Dufaux:2008dn,
    author = "Dufaux, Jean-Francois and Felder, Gary and Kofman, Lev and Navros, Olga",
    title = "{Gravity Waves from Tachyonic Preheating after Hybrid Inflation}",
    eprint = "0812.2917",
    archivePrefix = "arXiv",
    primaryClass = "astro-ph",
    reportNumber = "FTUAM-08-25, IFT-UAM-CSIC-08-90",
    doi = "10.1088/1475-7516/2009/03/001",
    journal = "JCAP",
    volume = "03",
    pages = "001",
    year = "2009"
}

@article{Bea:2021zol,
    author = "Bea, Yago and Casalderrey-Solana, Jorge and Giannakopoulos, Thanasis and Jansen, Aron and Krippendorf, Sven and Mateos, David and Sanchez-Garitaonandia, Mikel and Zilh\~ao, Miguel",
    title = "{Spinodal Gravitational Waves}",
    eprint = "2112.15478",
    archivePrefix = "arXiv",
    primaryClass = "hep-th",
    month = "12",
    year = "2021"
}

@article{Cook:2011hg,
    author = "Cook, Jessica L. and Sorbo, Lorenzo",
    title = "{Particle production during inflation and gravitational waves detectable by ground-based interferometers}",
    eprint = "1109.0022",
    archivePrefix = "arXiv",
    primaryClass = "astro-ph.CO",
    doi = "10.1103/PhysRevD.85.023534",
    journal = "Phys. Rev. D",
    volume = "85",
    pages = "023534",
    year = "2012",
    note = "[Erratum: Phys.Rev.D 86, 069901 (2012)]"
}

@article{Figueroa:2020rrl,
    author = "Figueroa, Daniel G. and Florio, Adrien and Torrenti, Francisco and Valkenburg, Wessel",
    title = "{The art of simulating the early Universe -- Part I}",
    eprint = "2006.15122",
    archivePrefix = "arXiv",
    primaryClass = "astro-ph.CO",
    doi = "10.1088/1475-7516/2021/04/035",
    journal = "JCAP",
    volume = "04",
    pages = "035",
    year = "2021"
}

@article{Chatrchyan:2023cmz,
    author = {Chatrchyan, Aleksandr and Er{\"o}ncel, Cem and Koschnitzke, Matthias and Servant, G{\'e}raldine},
    title = "{ALP dark matter with non-periodic potentials: parametric resonance, halo formation and gravitational signatures}",
    eprint = "2305.03756",
    archivePrefix = "arXiv",
    primaryClass = "hep-ph",
    reportNumber = "DESY-23-060",
    doi = "10.1088/1475-7516/2023/10/068",
    journal = "JCAP",
    volume = "10",
    pages = "068",
    year = "2023"
}

@article{Eroncel:2022efc,
    author = {Er{\"o}ncel, Cem and Servant, G{\'e}raldine},
    title = "{ALP dark matter mini-clusters from kinetic fragmentation}",
    eprint = "2207.10111",
    archivePrefix = "arXiv",
    primaryClass = "hep-ph",
    reportNumber = "DESY 22-115",
    doi = "10.1088/1475-7516/2023/01/009",
    journal = "JCAP",
    volume = "01",
    pages = "009",
    year = "2023"
}

@article{Eroncel:2022vjg,
    author = {Er{\"o}ncel, Cem and Sato, Ryosuke and Servant, Geraldine and S{\o}rensen, Philip},
    title = "{ALP dark matter from kinetic fragmentation: opening up the parameter window}",
    eprint = "2206.14259",
    archivePrefix = "arXiv",
    primaryClass = "hep-ph",
    reportNumber = "DESY 22-106, OU-HET-1148",
    doi = "10.1088/1475-7516/2022/10/053",
    journal = "JCAP",
    volume = "10",
    pages = "053",
    year = "2022"
}

@article{Freese:1990rb,
    author = "Freese, Katherine and Frieman, Joshua A. and Olinto, Angela V.",
    title = "{Natural inflation with pseudo - Nambu-Goldstone bosons}",
    reportNumber = "FERMILAB-PUB-90-177-A",
    doi = "10.1103/PhysRevLett.65.3233",
    journal = "Phys. Rev. Lett.",
    volume = "65",
    pages = "3233--3236",
    year = "1990"
}

@article{Daido:2017wwb,
    author = "Daido, Ryuji and Takahashi, Fuminobu and Yin, Wen",
    title = "{The ALP miracle: unified inflaton and dark matter}",
    eprint = "1702.03284",
    archivePrefix = "arXiv",
    primaryClass = "hep-ph",
    reportNumber = "TU-1039, IPMU17-0031",
    doi = "10.1088/1475-7516/2017/05/044",
    journal = "JCAP",
    volume = "05",
    pages = "044",
    year = "2017"
}

@article{Adshead:2015pva,
    author = "Adshead, Peter and Giblin, John T. and Scully, Timothy R. and Sfakianakis, Evangelos I.",
    title = "{Gauge-preheating and the end of axion inflation}",
    eprint = "1502.06506",
    archivePrefix = "arXiv",
    primaryClass = "astro-ph.CO",
    doi = "10.1088/1475-7516/2015/12/034",
    journal = "JCAP",
    volume = "12",
    pages = "034",
    year = "2015"
}

@article{Pajer:2013fsa,
    author = "Pajer, Enrico and Peloso, Marco",
    title = "{A review of Axion Inflation in the era of Planck}",
    eprint = "1305.3557",
    archivePrefix = "arXiv",
    primaryClass = "hep-th",
    doi = "10.1088/0264-9381/30/21/214002",
    journal = "Class. Quant. Grav.",
    volume = "30",
    pages = "214002",
    year = "2013"
}

@article{Kitajima:2021bjq,
    author = "Kitajima, Naoya and Nakagawa, Shota and Takahashi, Fuminobu",
    title = "{Nonthermally trapped inflation by tachyonic dark photon production}",
    eprint = "2111.06696",
    archivePrefix = "arXiv",
    primaryClass = "hep-ph",
    reportNumber = "TU-1131",
    doi = "10.1103/PhysRevD.105.103011",
    journal = "Phys. Rev. D",
    volume = "105",
    number = "10",
    pages = "103011",
    year = "2022"
}

@article{Domcke:2019qmm,
    author = "Domcke, Valerie and Ema, Yohei and Mukaida, Kyohei",
    title = "{Chiral Anomaly, Schwinger Effect, Euler-Heisenberg Lagrangian, and application to axion inflation}",
    eprint = "1910.01205",
    archivePrefix = "arXiv",
    primaryClass = "hep-ph",
    reportNumber = "DESY-19-166, DESY 19-166",
    doi = "10.1007/JHEP02(2020)055",
    journal = "JHEP",
    volume = "02",
    pages = "055",
    year = "2020"
}

@article{Takahashi:2021tff,
    author = "Takahashi, Fuminobu and Yin, Wen",
    title = "{Challenges for heavy QCD axion inflation}",
    eprint = "2105.10493",
    archivePrefix = "arXiv",
    primaryClass = "hep-ph",
    reportNumber = "TU-1124",
    doi = "10.1088/1475-7516/2021/10/057",
    journal = "JCAP",
    volume = "10",
    pages = "057",
    year = "2021"
}

@article{Salehian:2020dsf,
    author = "Salehian, Borna and Gorji, Mohammad Ali and Mukohyama, Shinji and Firouzjahi, Hassan",
    title = "{Analytic study of dark photon and gravitational wave production from axion}",
    eprint = "2007.08148",
    archivePrefix = "arXiv",
    primaryClass = "hep-ph",
    reportNumber = "YITP-20-90, IPMU20-0078",
    doi = "10.1007/JHEP05(2021)043",
    journal = "JHEP",
    volume = "05",
    pages = "043",
    year = "2021"
}

@article{Namba:2020kij,
    author = "Namba, Ryo and Suzuki, Motoo",
    title = "{Implications of Gravitational-wave Production from Dark Photon Resonance to Pulsar-timing Observations and Effective Number of Relativistic Species}",
    eprint = "2009.13909",
    archivePrefix = "arXiv",
    primaryClass = "astro-ph.CO",
    doi = "10.1103/PhysRevD.102.123527",
    journal = "Phys. Rev. D",
    volume = "102",
    pages = "123527",
    year = "2020"
}

@article{Kite:2020uix,
    author = "Kite, Thomas and Ravenni, Andrea and Patil, Subodh P. and Chluba, Jens",
    title = "{Bridging the gap: spectral distortions meet gravitational waves}",
    eprint = "2010.00040",
    archivePrefix = "arXiv",
    primaryClass = "astro-ph.CO",
    doi = "10.1093/mnras/stab1558",
    journal = "Mon. Not. Roy. Astron. Soc.",
    volume = "505",
    number = "3",
    pages = "4396--4405",
    year = "2021"
}

@article{Kitajima:2020rpm,
    author = "Kitajima, Naoya and Soda, Jiro and Urakawa, Yuko",
    title = "{Nano-Hz Gravitational-Wave Signature from Axion Dark Matter}",
    eprint = "2010.10990",
    archivePrefix = "arXiv",
    primaryClass = "astro-ph.CO",
    reportNumber = "KOBE-COSMO-20-16, TU-1111",
    doi = "10.1103/PhysRevLett.126.121301",
    journal = "Phys. Rev. Lett.",
    volume = "126",
    number = "12",
    pages = "121301",
    year = "2021"
}

@article{Banerjee:2021oeu,
    author = "Banerjee, Abhishek and Madge, Eric and Perez, Gilad and Ratzinger, Wolfram and Schwaller, Pedro",
    title = "{Gravitational wave echo of relaxion trapping}",
    eprint = "2105.12135",
    archivePrefix = "arXiv",
    primaryClass = "hep-ph",
    doi = "10.1103/PhysRevD.104.055026",
    journal = "Phys. Rev. D",
    volume = "104",
    number = "5",
    pages = "055026",
    year = "2021"
}

@article{Madge:2023dxc,
    author = "Madge, Eric and Morgante, Enrico and Puchades-Ib\'a\~nez, Cristina and Ramberg, Nicklas and Ratzinger, Wolfram and Schenk, Sebastian and Schwaller, Pedro",
    title = "{Primordial gravitational waves in the nano-Hertz regime and PTA data \textemdash{} towards solving the GW inverse problem}",
    eprint = "2306.14856",
    archivePrefix = "arXiv",
    primaryClass = "hep-ph",
    reportNumber = "MITP-23-029",
    doi = "10.1007/JHEP10(2023)171",
    journal = "JHEP",
    volume = "10",
    pages = "171",
    year = "2023"
}

@article{Su:2025nkl,
    author = "Su, Hong and Xu, Baoyu and Chen, Ju and Liu, Chang and Zhang, Yun-Long",
    title = "{Detectability of the chiral gravitational wave background from audible axions with the LISA-Taiji network}",
    eprint = "2503.20778",
    archivePrefix = "arXiv",
    primaryClass = "astro-ph.CO",
    month = "3",
    year = "2025"
}

@article{Anber:2012du,
    author = "Anber, Mohamed M. and Sorbo, Lorenzo",
    title = "{Non-Gaussianities and chiral gravitational waves in natural steep inflation}",
    eprint = "1203.5849",
    archivePrefix = "arXiv",
    primaryClass = "astro-ph.CO",
    doi = "10.1103/PhysRevD.85.123537",
    journal = "Phys. Rev. D",
    volume = "85",
    pages = "123537",
    year = "2012"
}

@article{Domcke:2016bkh,
    author = "Domcke, Valerie and Pieroni, Mauro and Bin\'etruy, Pierre",
    title = "{Primordial gravitational waves for universality classes of pseudoscalar inflation}",
    eprint = "1603.01287",
    archivePrefix = "arXiv",
    primaryClass = "astro-ph.CO",
    doi = "10.1088/1475-7516/2016/06/031",
    journal = "JCAP",
    volume = "06",
    pages = "031",
    year = "2016"
}

@article{Greene:1997fu,
    author = "Greene, Patrick B. and Kofman, Lev and Linde, Andrei D. and Starobinsky, Alexei A.",
    title = "{Structure of resonance in preheating after inflation}",
    eprint = "hep-ph/9705347",
    archivePrefix = "arXiv",
    reportNumber = "SU-ITP-97-19, IFA-97-29",
    doi = "10.1103/PhysRevD.56.6175",
    journal = "Phys. Rev. D",
    volume = "56",
    pages = "6175--6192",
    year = "1997"
}

@article{Kofman:1994rk,
    author = "Kofman, Lev and Linde, Andrei D. and Starobinsky, Alexei A.",
    title = "{Reheating after inflation}",
    eprint = "hep-th/9405187",
    archivePrefix = "arXiv",
    reportNumber = "UH-IFA-94-35, SU-ITP-94-13, YITP-U-94-15",
    doi = "10.1103/PhysRevLett.73.3195",
    journal = "Phys. Rev. Lett.",
    volume = "73",
    pages = "3195--3198",
    year = "1994"
}

@article{Maleknejad:2016qjz,
    author = "Maleknejad, Azadeh",
    title = "{Axion Inflation with an SU(2) Gauge Field: Detectable Chiral Gravity Waves}",
    eprint = "1604.03327",
    archivePrefix = "arXiv",
    primaryClass = "hep-ph",
    doi = "10.1007/JHEP07(2016)104",
    journal = "JHEP",
    volume = "07",
    pages = "104",
    year = "2016"
}

@article{Cuissa:2018oiw,
    author = "Cuissa, Jose Roberto Canivete and Figueroa, Daniel G.",
    title = "{Lattice formulation of axion inflation. Application to preheating}",
    eprint = "1812.03132",
    archivePrefix = "arXiv",
    primaryClass = "astro-ph.CO",
    doi = "10.1088/1475-7516/2019/06/002",
    journal = "JCAP",
    volume = "06",
    pages = "002",
    year = "2019"
}

@article{Morgante:2022zvc,
    author = "Morgante, Enrico and Ramberg, Nicklas and Schwaller, Pedro",
    title = "{Gravitational waves from dark SU(3) Yang-Mills theory}",
    eprint = "2210.11821",
    archivePrefix = "arXiv",
    primaryClass = "hep-ph",
    doi = "10.1103/PhysRevD.107.036010",
    journal = "Phys. Rev. D",
    volume = "107",
    number = "3",
    pages = "036010",
    year = "2023"
}

@article{Morgante:2021bks,
    author = "Morgante, Enrico and Ratzinger, Wolfram and Sato, Ryosuke and Stefanek, Ben A.",
    title = "{Axion fragmentation on the lattice}",
    eprint = "2109.13823",
    archivePrefix = "arXiv",
    primaryClass = "hep-ph",
    reportNumber = "MITP-21-045",
    doi = "10.1007/JHEP12(2021)037",
    journal = "JHEP",
    volume = "12",
    pages = "037",
    year = "2021"
}

@article{Kierkla:2025vwp,
    author = "Kierkla, Maciej and Ramberg, Nicklas and Schicho, Philipp and Schmitt, Daniel",
    title = "{Theoretical uncertainties for primordial black holes from cosmological phase transitions}",
    eprint = "2506.15496",
    archivePrefix = "arXiv",
    primaryClass = "hep-ph",
    reportNumber = "SISSA 07/2025/FISI",
    month = "6",
    year = "2025"
}

@article{Kierkla:2023von,
    author = "Kierkla, Maciej and Swiezewska, Bogumila and Tenkanen, Tuomas V. I. and van de Vis, Jorinde",
    title = "{Gravitational waves from supercooled phase transitions: dimensional transmutation meets dimensional reduction}",
    eprint = "2312.12413",
    archivePrefix = "arXiv",
    primaryClass = "hep-ph",
    doi = "10.1007/JHEP02(2024)234",
    journal = "JHEP",
    volume = "02",
    pages = "234",
    year = "2024"
}

@article{Kierkla:2025qyz,
    author = "Kierkla, Maciej and Schicho, Philipp and Swiezewska, Bogumila and Tenkanen, Tuomas V. I. and van de Vis, Jorinde",
    title = "{Finite-temperature bubble nucleation with shifting scale hierarchies}",
    eprint = "2503.13597",
    archivePrefix = "arXiv",
    primaryClass = "hep-ph",
    reportNumber = "CERN-TH-2025-046, HIP-2024-27/TH",
    doi = "10.1007/JHEP07(2025)153",
    journal = "JHEP",
    volume = "07",
    pages = "153",
    year = "2025"
}

@article{Kierkla:2022odc,
    author = "Kierkla, Maciej and Karam, Alexandros and Swiezewska, Bogumila",
    title = "{Conformal model for gravitational waves and dark matter: a status update}",
    eprint = "2210.07075",
    archivePrefix = "arXiv",
    primaryClass = "astro-ph.CO",
    doi = "10.1007/JHEP03(2023)007",
    journal = "JHEP",
    volume = "03",
    pages = "007",
    year = "2023"
}

@article{Sagunski:2023ynd,
    author = "Sagunski, Laura and Schicho, Philipp and Schmitt, Daniel",
    title = "{Supercool exit: Gravitational waves from QCD-triggered conformal symmetry breaking}",
    eprint = "2303.02450",
    archivePrefix = "arXiv",
    primaryClass = "hep-ph",
    doi = "10.1103/PhysRevD.107.123512",
    journal = "Phys. Rev. D",
    volume = "107",
    number = "12",
    pages = "123512",
    year = "2023"
}

@article{Caprini:2019egz,
    author = "Caprini, Chiara and others",
    title = "{Detecting gravitational waves from cosmological phase transitions with LISA: an update}",
    eprint = "1910.13125",
    archivePrefix = "arXiv",
    primaryClass = "astro-ph.CO",
    reportNumber = "DESY-19-159, IPPP/19/27, HIP-2019-14/TH, MITP/19-066, IFT-UAM/CSIC-19-139",
    doi = "10.1088/1475-7516/2020/03/024",
    journal = "JCAP",
    volume = "03",
    pages = "024",
    year = "2020"
}

@article{Schwaller:2015tja,
    author = "Schwaller, Pedro",
    title = "{Gravitational Waves from a Dark Phase Transition}",
    eprint = "1504.07263",
    archivePrefix = "arXiv",
    primaryClass = "hep-ph",
    reportNumber = "CERN-PH-TH-2015-093",
    doi = "10.1103/PhysRevLett.115.181101",
    journal = "Phys. Rev. Lett.",
    volume = "115",
    number = "18",
    pages = "181101",
    year = "2015"
}

@article{Croon:2020cgk,
    author = "Croon, Djuna and Gould, Oliver and Schicho, Philipp and Tenkanen, Tuomas V. I. and White, Graham",
    title = "{Theoretical uncertainties for cosmological first-order phase transitions}",
    eprint = "2009.10080",
    archivePrefix = "arXiv",
    primaryClass = "hep-ph",
    reportNumber = "HIP-2020-26/TH",
    doi = "10.1007/JHEP04(2021)055",
    journal = "JHEP",
    volume = "04",
    pages = "055",
    year = "2021"
}

@article{Hindmarsh:2013xza,
    author = "Hindmarsh, Mark and Huber, Stephan J. and Rummukainen, Kari and Weir, David J.",
    title = "{Gravitational waves from the sound of a first order phase transition}",
    eprint = "1304.2433",
    archivePrefix = "arXiv",
    primaryClass = "hep-ph",
    reportNumber = "HIP-2013-07-TH",
    doi = "10.1103/PhysRevLett.112.041301",
    journal = "Phys. Rev. Lett.",
    volume = "112",
    pages = "041301",
    year = "2014"
}

@article{Greene:1998pb,
    author = "Greene, Patrick B. and Kofman, Lev and Starobinsky, Alexei A.",
    title = "{Sine-Gordon parametric resonance}",
    eprint = "hep-ph/9808477",
    archivePrefix = "arXiv",
    reportNumber = "CITA-98-33",
    doi = "10.1016/S0550-3213(99)00018-8",
    journal = "Nucl. Phys. B",
    volume = "543",
    pages = "423--443",
    year = "1999"
}

@article{Nakagawa:2020zjr,
    author = "Nakagawa, Shota and Takahashi, Fuminobu and Yamada, Masaki",
    title = "{Trapping Effect for QCD Axion Dark Matter}",
    eprint = "2012.13592",
    archivePrefix = "arXiv",
    primaryClass = "hep-ph",
    reportNumber = "TU-1116, IPMU20-0132",
    doi = "10.1088/1475-7516/2021/05/062",
    journal = "JCAP",
    volume = "05",
    pages = "062",
    year = "2021"
}

@book{McLachlan:1947:TA,
    author = {McLachlan, N. W.},
     title = {Theory and Application of {M}athieu Functions},
 publisher = {Clarendon Press},
   address = {Oxford},
      year = {1947},
     pages = {xii+401},
      note = {Corrected republication by Dover Publications, Inc., New York, 1964.},
   mrclass = {33.0X},
  mrnumber = {MR0021158 (9,31b)},
 mrreviewer = {M. J. O. Strutt},
     zblno = {0029.02901}}

@article{McAllister:2014mpa,
    author = "McAllister, Liam and Silverstein, Eva and Westphal, Alexander and Wrase, Timm",
    title = "{The Powers of Monodromy}",
    eprint = "1405.3652",
    archivePrefix = "arXiv",
    primaryClass = "hep-th",
    reportNumber = "SU/ITP-14/13, SLAC-PUB-15962, DESY-14-078, SU-ITP-14-13",
    doi = "10.1007/JHEP09(2014)123",
    journal = "JHEP",
    volume = "09",
    pages = "123",
    year = "2014"
}

@article{Silverstein:2008sg,
    author = "Silverstein, Eva and Westphal, Alexander",
    title = "{Monodromy in the CMB: Gravity Waves and String Inflation}",
    eprint = "0803.3085",
    archivePrefix = "arXiv",
    primaryClass = "hep-th",
    reportNumber = "SU-ITP-08-07, SLAC-PUB-13183",
    doi = "10.1103/PhysRevD.78.106003",
    journal = "Phys. Rev. D",
    volume = "78",
    pages = "106003",
    year = "2008"
}

@article{McAllister:2008hb,
    author = "McAllister, Liam and Silverstein, Eva and Westphal, Alexander",
    title = "{Gravity Waves and Linear Inflation from Axion Monodromy}",
    eprint = "0808.0706",
    archivePrefix = "arXiv",
    primaryClass = "hep-th",
    reportNumber = "SLAC-PUB-13357, SU-ITP-08-15",
    doi = "10.1103/PhysRevD.82.046003",
    journal = "Phys. Rev. D",
    volume = "82",
    pages = "046003",
    year = "2010"
}

@article{Marchesano:2014mla,
    author = "Marchesano, Fernando and Shiu, Gary and Uranga, Angel M.",
    title = "{F-term Axion Monodromy Inflation}",
    eprint = "1404.3040",
    archivePrefix = "arXiv",
    primaryClass = "hep-th",
    reportNumber = "IFT-UAM-CSIC-14-032, MAD-TH-04-01",
    doi = "10.1007/JHEP09(2014)184",
    journal = "JHEP",
    volume = "09",
    pages = "184",
    year = "2014"
}

@article{Blumenhagen:2014gta,
    author = "Blumenhagen, Ralph and Plauschinn, Erik",
    title = "{Towards Universal Axion Inflation and Reheating in String Theory}",
    eprint = "1404.3542",
    archivePrefix = "arXiv",
    primaryClass = "hep-th",
    doi = "10.1016/j.physletb.2014.08.007",
    journal = "Phys. Lett. B",
    volume = "736",
    pages = "482--487",
    year = "2014"
}

@article{Hebecker:2014eua,
    author = "Hebecker, Arthur and Kraus, Sebastian C. and Witkowski, Lukas T.",
    title = "{D7-Brane Chaotic Inflation}",
    eprint = "1404.3711",
    archivePrefix = "arXiv",
    primaryClass = "hep-th",
    doi = "10.1016/j.physletb.2014.08.028",
    journal = "Phys. Lett. B",
    volume = "737",
    pages = "16--22",
    year = "2014"
}
%%%%%%%%%%%%%%%%%%%%%%%%%%%%%%%%%%%%%%%%

\end{document}